\documentclass{ar-1col-S2O}
\usepackage[comma]{natbib}

\usepackage{amsmath} 

\usepackage{graphicx}
\usepackage{subcaption} 
\def\bm#1{\mathbf{#1}}

\def\balpha{\mbox{\boldmath$\alpha$}}
\def\bbeta{\mbox{\boldmath$\beta$}}
\def\bgamma{\mbox{\boldmath$\gamma$}}
\def\bdelta{\mbox{\boldmath$\delta$}}
\def\bvarepsilon{\mbox{\boldmath$\varepsilon$}}
\def\bepsilon{\mbox{\boldmath$\epsilon$}}
\def\btheta{\mbox{\boldmath$\theta$}}
\def\bkappa{\mbox{\boldmath$\kappa$}}
\def\bmu{\mbox{\boldmath$\mu$}}
\def\bpi{\mbox{\boldmath$\pi$}}
\def\bomega{\mbox{\boldmath$\omega$}}

\def\bIq{\bm{I}_q}
\def\bQ{\bm{Q}}
\def\bD{\bm{D}}
\def\bG{\bm{G}}

\def\bA{\bm{A}}
\def\bB{\bm{B}}
\def\bC{\bm{C}}
\def\bJ{\bm{J}}
\def\bM{\bm{M}}

\def\bx{\bm{x}}

\def\bz{\bm{z}}

\def\df{{\rm df}}

\def\Gam{{\rm Gam}}

\def\narr{{\rm narr}}
\def\wide{{\rm wide}}
\def\sqb{{\rm sqb}}

\def\beq{\begin{eqnarray}}
\def\eeq{\end{eqnarray}}

\def\beqn{\begin{eqnarray*}}  
\def\eeqn{\end{eqnarray*}}

\def\dellone{\hbox{$\dell\mu\over \dell\boldsymbol\theta$}}
\def\delltwo{\hbox{$\dell\mu\over \dell\boldsymbol\gamma$}}
\def\para{{\rm para}}
\def\nonpara{{\rm nonp}}

\def\E{{\rm E}}

\def\dd{{\rm d}}
\def\N{{\rm N}}
\def\Pr{{\rm P}}
\def\pr{{\rm pr}}

\def\quadandquad{\quad \hbox{and} \quad}
\def\arr{\rightarrow}
\def\hatt{\widehat}
\def\tilda{\widetilde}
\def\sumin{\sum_{i=1}^n}
\def\sumjk{\sum_{j=1}^k}
\def\eps{\varepsilon}
\def\half{\hbox{$1\over2$}}
\def\rootn{\sqrt{n}}
\def\data{{\rm data}}
\def\midd{\,|\,}
\def\tr{{\top}}
\def\dell{\partial}

\def\true{{\rm true}}
\def\med{{\rm med}}
\def\AIC{{\rm AIC}}
\def\FIC{{\rm FIC}}
\def\AFIC{{\rm AFIC}}
\def\Tr{{\rm Tr}}
\def\mse{{\rm mse}}

\def\KL{{\rm KL}}

\def\fic{{\rm fic}}
\def\afic{{\rm afic}}
\def\fric{{\rm fric}}
\def\nonpara{{\rm np}}

\begin{document}

\markboth{Claeskens and Hjort}{Focused information criteria}

\title{Focused Information Criteria}

\author{Gerda Claeskens$^1$ and Nils Lid Hjort$^2$
\affil{$^1$ORStat and Leuven Statistics Research Center,
    KU Leuven, 3000 Leuven, Belgium; gerda.claeskens@kuleuven.be}
\affil{$^2$Department of Mathematics, University of Oslo,
  Norway; nils@math.uio.no}
}

\begin{abstract}
  The focused information criterion is used to make a choice
  among several statistical models, or among several
  variables to include in a model. Different from other
  such information criteria, the focused information criterion
  is constructed to select the best model for a given interest
  quantity, the focus of the research question.
  Different such focus parameters may lead to
  different selected models, each one best for the corresponding
  focus. What is `best' is defined by a risk function,
  often the mean squared error. Other risks can be considered
  too for focused selection. Selections by the focused information
  criterion include using parametric (generalized) linear models,
  non- and semiparametric models, quantile regression models,
  graphical models, models for survival data,
  for longitudinal data, time series models, and many more.
  Extensions of the basic version include versions
  for high-dimensional data, regularized estimation,
  and Bayesian methods.
\end{abstract}

\maketitle

\tableofcontents

\section{INTRODUCTION}
\label{section:intro}

Building good models for one's data, quite often involving
searching through candidate models before finding an adequate one,
is of course a major theme of theoretical and applied statistics.
The relative ease with which even a high number of candidate
models can be worked through (`I just ran two million regressions'
is the pertinent title of \citet{SalaiMartin97}) highlights the need
to have automated well-founded model comparison and model ranking
methods. Such have been worked with since the 1970ies and 1980ies,
with the most well-known versatile methods AIC and BIC
(the Akaike and the Bayesian Information Criterion)
in extensive use; see \citet{ClaeskensHjort08} and
the review article \citet{Claeskens16}.
The present article reviews the differently spirited FIC,
the Focused Information Criterion, which takes into account
the leading questions of interest when sorting through
and ranking candidate models.

Data collection and model building is indeed often driven
by underlying focused questions; one wishes to find
the best model related to such. Some instances are
pointed to here, with yet others discussed later in our article:
In biology, the population size estimation of whale sharks
in a certain region might be of special interest \citep{Farcomeni17};
in \citet{CunenHjort18, CunenWalloeHjort20, CunenWalloeKonishiHjort21},
the marine biology focus is the body condition of minke whales,
monitored over time.
With functional magnetic resonance imaging (fMRI) data
researchers may focus on estimating the mean signal
in specific brain regions of interest
\citep{PircalabeluClaeskensJahfariWaldorp15,
PircalabeluClaeskensWaldorp15}.
In time series analysis in an economical context one wishes
to select the order and the innovation density of
autoregressive processes \citep{PandhareRamanathan2020}.
For \citet{WangHH2019} the focus lies on the claim severity
distribution for actuarial data.
\citet{BartoliccuLupparelli08} focus on the population size
using capture-recapture models.
Using quantile regression models one wishes to estimate
the minimum effective dose in phase II clinical trials
\citep{BehlClaeskensDette2014}.
In \cite{YangLiuLiang2015, HelltonHjort18}
the aim is that of obtaining good personalized prediction
for individual patients in a medical setting.
In various medical studies one wishes to estimate the survival
probability for a given patient when other medical information is available
\citep{HjortClaeskens06, Hjort08Vonta, HjortStoltenberg23}.
For \citet{Haug19}, the task is to select dynamically evolving
Markov chain models for the level of conflict or war,
year by year, aiming for the most precise probability
estimates for further escalation.

For each of the situations pointed to here there would be
multiple candidate models, to be sorted through and ranked;
these could reflect different selections of covariates,
potential interaction factors,
additional parameters related to skewnesses of error distributions, etc.
What these setups further have in common is that one wishes
to estimate a certain chief quantity, with a clear
interpretation across candidate models.
The idea of using the focus, the quantity of interest, to select a model, makes sense because one model cannot be best for estimation of everything. 
When performing focused selection, the research question is translated into a focus parameter, which is an estimable parameter that is interpretable in the available set of models to choose from. Starting from the focus parameter, the FIC estimates for each model the risk involved to estimate the focus. The model with the minimum such estimated risk is the model selected by the FIC.

A generic formalized version of these main ideas is as follows.
There is a focus parameter $\mu$, with clear interpretation
across candidate models $M_1,\ldots,M_k$. Using model $M_j$
leads to the estimator $\hatt\mu_j$, in terms of that model's
estimated parameters. Though there are other variants,
the quality of this outcome
of using $M_j$ is defined as the mean squared error (mse)
\beq
\mse_j=\E_\true[(\hatt\mu_j-\mu_\true)^2]
=\sqb_j + v_j,
\label{eq:msebasic}
\eeq
say, in terms of the squared bias $\sqb_j=b_j^2$ of the estimator
$\widehat{\mu}_j$ and its variance $v_j$.
These are defined in terms of the true underlying model
having generated the data. The FIC idea is to assess
and estimate the squared bias and the variance,
perhaps via good approximations, and in ways depending
on model assumptions for the true model, and then to arrive at
\beq
\FIC_j=\hatt{\mse_j}=\hatt\sqb_j + \hatt v_j.
\label{eq:ficbasic}
\eeq
This gives a FIC ranking, from the best models with low
scores to the worst with higher scores. 
The root-FIC scores $\FIC_j^{1/2}$ are interpretable as
estimated precision on the scale of $\mu$ itself.
We note here that if $\hatt b_j$ is a good estimator
of the bias itself, the square $\hatt b_j^2$
will overshoot $\sqb_j$, its mean being approximately
$b_j^2+\tau_j$, say, with $\tau_j$ the variance of $\widehat{b}_j$.
The $\sqb_j$ estimator ought therefore to be of the
type $\hatt b_j^2-\hatt\tau_j$, typically truncated
to zero to avoid negative values.
It is also to be noted that different foci might easily
lead to different rankings and different best models,
unlike most other selection schemes in mainstream use.

Though the FIC formula (\ref{eq:ficbasic}) looks innocently generic,
the ingredients will take different forms in different
setups, constructing the appropriate versions of
$\max(\hatt b_j^2-\hatt\tau_j,0)+\hatt v_j$.
These will depend not only on the list of candidate
models, and the focus $\mu$, but also, crucially,
on what is assumed about the underlying true model.
Carrying out FIC analysis and model ranking in practice
can be relatively easy or relative complex, depending
on these factors; thus its implementation might typically
take more efforts than the familiar AIC and BIC criteria,
for which merely the log-likelihood maxima are required.
The gain of using FIC lies in precision: one selects the
best estimator for the studied focus, in contrast to AIC, BIC
and similar criteria, which do not at all take the purpose
of the selection into account.

In Section \ref{section:ficA} we give broadly applicable
formulae for FIC in the setting of general regression models.
There is a narrow model of dimension $p$ and a wide model
requiring $q$ more parameters, with all $2^q$ subsets
of these being associated with possible candidate submodels.
This fits the typical situation with `protected'
covariates $x_1,\ldots,x_p$, those being present
in all models considered, and `open' covariates
$z_1,\ldots,z_q$, and subset models corresponding including
some of these while excluding the others.
The operative assumption is that the distance
from the narrow to the wide model is moderate.
The setup of Section \ref{section:ficB} is different,
where parametric candidate models are compared
with a fixed wide model, perhaps nonparametrically defined.
Section \ref{section:aficandmore}
broadens the horizon, with average-FIC procedures
and their connections to model testing,
Section \ref{section:fic-finetuning} gives 
strategies for using FIC to select fine-tuning parameters
in different settings and to select sometimes better estimators than
the maximum likelihood one.
Issues of post-selection inference are discussed
in Section \ref{section:postselection}.
Variations of the FIC, e.g.~with different loss functions,
are worked with in Section \ref{section:othermethods}.
Then Section \ref{section:recent} offers pointers
to recent methodological developments, partly
driven by an increasing list of application domains.
Some concluding comments are in Section \ref{section:concluding}.
The separately available online supplement contains
additional details, pertaining to the wide model FIC,
influence functions, and FIC for other loss functions. 

\section{FIC IN THE LOCAL NEIGHBORHOOD MODELS FRAMEWORK}
\label{section:ficA}

It is clear from the general setup above that the
formulae for the mse in eq.~(\ref{eq:msebasic})
and their estimators in eq.~(\ref{eq:ficbasic}) will pan
out differently in different frameworks.
This is not merely since estimands $\mu$ and
estimators $\hatt\mu$ are of different types,
but because the modelling assumptions about
the true underlying data generating process matters
crucially. In the present section we are able
to find quite general formulae, valid for all smooth
parametric models, in a certain $O(1/\rootn)$
modelling framework;
in Section \ref{section:ficB} somewhat different formulae
are exhibited, for the different setup of having
a fixed wide data-generating model.

\subsection{Defining the focus and the FIC value}
\label{subsection:ficA-deffocus}

In a parametric regression model, the true density (or probability mass function) of a response variable $Y$ depends on some parameters. We distinguish between `protected' parameters, denoted by \btheta, that are present in all of the models and are not subject to selection, and the parameter vector \bgamma~that is the topic of the model search; some of its components may appear in the selected models, the components that are not, are called `non-selected'.

For a spelled-out fairly typical situation of this sort,
consider a normal linear model with responses
$Y_i=\bm{x}_i^\tr\bbeta+\bm{z}_i^\tr\bgamma+\eps_i$
for $i=1,\ldots,n$, with $\bm{x}_i$ the protected covariates of length $p$
and $\bm{z}_i$ the open covariates of length $q$,
and with the $\eps_i$ being i.i.d.~$\N(0,\sigma^2)$.
The narrow model has the $p+1$ parameters $(\bbeta^\top,\sigma)$,
with none of the $z_{i,j}$ included, whereas the wide
model has the $p+q+1$ parameters $(\bbeta^\top,\sigma,\bgamma^\top)$,
with all of $z_{i,1},\ldots,z_{i,q}$ on board.
Focus parameters could be
(i) the mean response
$\mu_1=\E\,(Y\midd \bm{x}_0,\bm{z}_0)=\bm{x}_0^\tr\bbeta+\bm{z}_0^\tr\bgamma$, for a
fixed or perhaps new position $(\bm{x}_0,\bm{z}_0)$ in the covariate space;
(ii) the 0.90 quantile
$\mu_2=\bm{x}_0^\tr\bbeta+\bm{z}_0^\tr\bgamma+\sigma\,z_{0.90}$; or
(iii) the probability
$\mu_3=\Pr(Y\ge y_0\midd \bm{x}_0,\bm{z}_0)$ of such a new $Y$
exceeding a relevant threshold $y_0$.
For each subset $S\subset\{1,\ldots,q\}$
the corresponding submodel can be used,
with $z_{i,j}$ in the model for $j\in S$ but not with $j\notin S$,
leading to estimators $\hatt\mu_S$.

With each estimator $\hatt\mu_S$ of the focus $\mu$
there is an associated bias $b_S$ and variance $v_S$, needing
a careful definition 
before we can properly construct the ensuing FIC scores,
say $\FIC_S$, as per eq.~(\ref{eq:ficbasic}). The bias $b_S$
in particular depends crucially on what is assumed
to be an adequate representation of the true model.
The typical option here is to assume that the full wide
model, containing all of $\bbeta,\sigma,\bgamma$, is true.

In order to approximate and estimate the $\mse$ of formula (\ref{eq:msebasic}),
to arrive at FIC 
as in eq.~(\ref{eq:ficbasic}),
assumptions are needed. In \cite{ClaeskensHjort03} we used
a locally misspecified model in which
$f_\true(y)=f(y;\bm{x},\btheta_0,\bgamma_0+\bdelta/\rootn)$,
here spelled out for regression contexts.
In this scenario the squared bias and variance of the
maximum likelihood (ML) estimators are of the same order
and neither one dominates.
The narrow model, of dimension $p$, corresponds to $\bgamma=\bgamma_0$,
a known value, like zero; the wide model has dimension $p+q$.
The true focus parameter is $\mu_\true=\mu(\btheta_0,\bgamma_0+\bdelta/\rootn)$,
with submodel estimators of the form
$\hatt\mu_S=\mu(\widehat{\btheta}_S,\widehat{\bgamma}_S,\bgamma_{0,S^c})$,
with $(\widehat{\btheta}_S,\widehat{\bgamma}_S)$ ML
estimators in the $S$ subset model, i.e.~the maximizers of
that model's log-likelihood function
$ 
\ell_{n,S}(\btheta,\bgamma_S)=\sumin \log f(y_i;\bm{x}_i,\btheta,
   \bgamma_S,{\bgamma}_{0,S^c}).
$ 
Under this true density function, one can compute for each focus estimator
its bias and variance. Exact calculations are available
in normal linear models, but require approximations
via Taylor series expansions and Lindeberg approximate normality
theorems in other modelling setups.

Such broadly applicable results have been obtained
in \citet{ClaeskensHjort03, ClaeskensHjort08},
to be described now.
They do involve a certain amount of book-keeping matrices,
related to biases and back-and-forth projections
for as many as $2^q$ correlated limiting normals.
Define first the $(p+q)\times(p+q)$ Fisher information matrix
and its inverse, for the model $f(y;\bm{x},\btheta,\bgamma)$,
evaluated at the narrow model parameters $(\btheta_0,\bgamma_0)$,
\beq
\bm{J}_\wide=\begin{pmatrix}
   \bm{J}_{00}, &\bm{J}_{01} \\
   \bm{J}_{10}, &\bm{J}_{11} \\ \end{pmatrix}
\quadandquad
\bm{J}_\wide^{-1}=\begin{pmatrix}
   \bm{J}^{00}, &\bm{J}^{01} \\
   \bm{J}^{10}, &\bm{J}^{11} \\ \end{pmatrix}.
\label{eq:Jwide}
\eeq
A crucial component is the variance related to the ML estimator
of $\bgamma$, the $q\times q$ matrix
\beq
\bm{Q}=\bm{J}^{11} = ({\bJ}_{11}-{\bJ}_{10} {\bJ}_{00}^{-1} {\bJ}_{01})^{-1}.
\label{eq:hereisQ}
\eeq
Define next, with partial derivatives evaluated at the narrow model,
\beq
\tau_0^2=\dellone^\tr \bm{J}_{00}^{-1}\dellone
\quadandquad
\bomega=\bm{J}_{10}\bm{J}_{00}^{-1}\dellone - \delltwo.
\label{eq:tau0andomega}
\eeq
We finally need projection matrices $\bpi_S$, taking $\bm{v}=(v_1,\ldots,v_q)^\top$
to the subvector $\bm{v}_S$ of size $|S|$ (the cardinality of $S$) containing only the components with $j\in S$, and
\beqn
\bm{G}_S=\bpi_S^\tr\bm{Q}_S\bpi_S\bm{Q}^{-1}
= \bpi_S^\tr(\bpi_S\bm{Q}^{-1}\bpi_S^\tr)^{-1}\bpi_S \bm{Q}^{-1}.
\eeqn
These are $q\times q$ matrices with the trace (sum of the diagonal
elements of the matrix) $\Tr(\bm{G}_S)=|S|$.
For the narrow and wide models, the $G_S$ is equal to
$\bm{0}$ and $\bIq$, respectively.
If $\bm{Q}$ is a diagonal matrix with $(\kappa_1,\ldots,\kappa_q)$
on its diagonal,
then $\bm{G}_S$ is the diagonal matrix with $\kappa_j$ for $j\in S$
and zeroes elsewhere on the diagonal.
The relevant and useful Master Theorem I now says that
under suitable regularity conditions,
\beq
\rootn(\widehat{\mu}_S-\mu_\true)\arr_d
   \Lambda_S=\Lambda_0 + \bomega^\tr(\bdelta - \bm{G}_S\bm{D}),
   \label{eq:master1}
\eeq
with joint convergence for all $2^q$ components.
This is a limit distribution representation for all the
submodel estimators $\widehat{\mu}_S$, in terms of
a basic common $\Lambda_0\sim\N({0},\tau_0^2)$
and an independent $\bm{D}\sim\N_q(\bdelta,\bm{Q})$.

In the limit experiment, defined in terms of the independent
normals $\Lambda_0$ and $\bm{D}$, all quantities can be
consistently estimated from data, apart from $\bdelta$
in the $O(\bdelta/\rootn)$ setup, and for which we merely
observe $\bm{D}$, which is also the limit distribution for
$\bm{D}_n=\rootn(\hatt{\bgamma}_\wide-\bgamma_0)$.
The limiting mse for $\rootn\hatt{\mu}_S$, as an estimator
of $\rootn\mu_\true$, is
\beqn
\mse_S=\tau_0^2+\bomega^\tr \bm{G}_S\bm{Q}\bm{G}_S^\tr\bomega
   +\{\bomega^\tr(\bm{I}_q-\bm{G}_S)\bdelta\}^2,
\eeqn
with $\bm{I}_q$ the $q\times q$ identity matrix.
Noting the $\E\,(\bm{DD}^\tr)=\bdelta\bdelta^\tr+\bm{Q}$,
an unbiased estimator for $\mse_S$ is
\beqn
\hatt\mse_S=\tau_0^2+\bomega^\tr \bm{G}_S\bm{Q}\bm{G}_S^\tr\bomega
   +\bomega^\tr(\bm{I}_q-\bG_S)(\bm{DD}^\tr-\bm{Q})(\bm{I}_q-\bG_S) ^\tr\bomega.
\eeqn
The `real FIC' now emerges by plugging in estimators
for the relevant quantities, i.e.
\beqn
\fic_S=\hatt{\tau}_0^2+\hatt{\bomega}^\tr \hatt{\bm{G}}_S\hatt{\bQ}
   \hatt{\bm{G}}_S^\tr\hatt{\bomega}
+
\max\{\hatt{\bomega}^\tr(\bm{I}_q-\hatt{\bG}_S)(\bm{D}_n\bm{D}_n^\tr-\hatt{\bQ})
   (\bm{I}_q-\hatt{\bG}_S) ^\tr\hatt{\bomega},0\}, 
\eeqn
where with the added truncation
the estimated squared bias is always nonnegative.

The normalized Fisher information matrix $\bm{J}_\wide$ of eq.~(\ref{eq:Jwide}) is estimated first,
 e.g.,
$\hat{\bm{J}}_\wide= -n^{-1}\dell^2\ell_{n,\wide}(\balpha)/\dell\balpha\,\dell\balpha^\tr$, 
writing $\balpha=(\btheta,\bgamma)$.
From this we obtain estimates of $\bm{Q}$ and $\bm{G}_S$.
For $\bomega$ one needs partial derivatives of $\mu(\btheta,\bgamma)$,
computed at the ML position, via exact formulae in
the simpler models or via numerical methods.
We note that $\fic_S$ has emerged by estimating
$\mse_S$ on the $\rootn(\hatt\mu_S-\mu_\true)$ scale,
so for tables and plots we use $\fic_S/n$,
or $\fic_S^{1/2}/\rootn$ for the root-fic scores;
these are the values plotted in \textbf{Figures \ref{figure:eyes}}
and \textbf{\ref{figure:egypt}}.

\subsection{Illustration: Diabetic retinopathy study}
\label{subsection:klein2008}

In the broad study \citet{Klein08}, the aim was to examine
the 25-year cumulative progression and regression of
diabetic retinopathy, in the light of
various risk factors. It involved following insulin-taking persons
living in Wisconsin with type 1 diabetes diagnosed before age 30.
The main outcome $y$ is an indicator for moderate
to severe nonproliferate retinopathy, or proliferate retinopathy,
for one or both eyes; out of $n=691$ individuals
with no missing covariates, there are 134 with $y=1$
and $557$ with $y=0$. We have organized data
into rows of $(x_1,x_2,z_1,z_2,z_3,z_4,z_5,y)$, with
$x_1$, duration since diagnosis;
$x_2$, indicator for presence of macular edema in one or both eyes;
$z_1$, glycosylated hemoglobin level;
$z_2$, body-mass index bmi;
$z_3$, pulse rate;
$z_4$, gender (1 for male, 0 for female);
$z_5$, indicator for presence of urine protein.
We treat the intercept and $x_1,x_2$ as protected,
i.e.~they are included in each candidate model, 
whereas $z_1,\ldots,z_5$ are open, i.e.~can be included
or excluded in submodels, for different purposes. This
leads to searching through $2^5=32$ logistic regression models,
submodels of the wide model of the form
\beqn
p(\bm{x},\bm{z})
=\Pr(Y=1\midd \bm{x},\bm{z})
=H(\beta_0+\beta_1x_1+\beta_2x_2+
   \gamma_1z_1+\gamma_2z_2+\gamma_3z_3+\gamma_4z_4+\gamma_5z_5),
\eeqn
with $H(u)=\exp(u)/\{1+\exp(u)\}$ the logistic transform.

\begin{figure}[h]
  \includegraphics[width=0.9\textwidth,trim={3cm 0.6cm 3cm 2cm},clip]{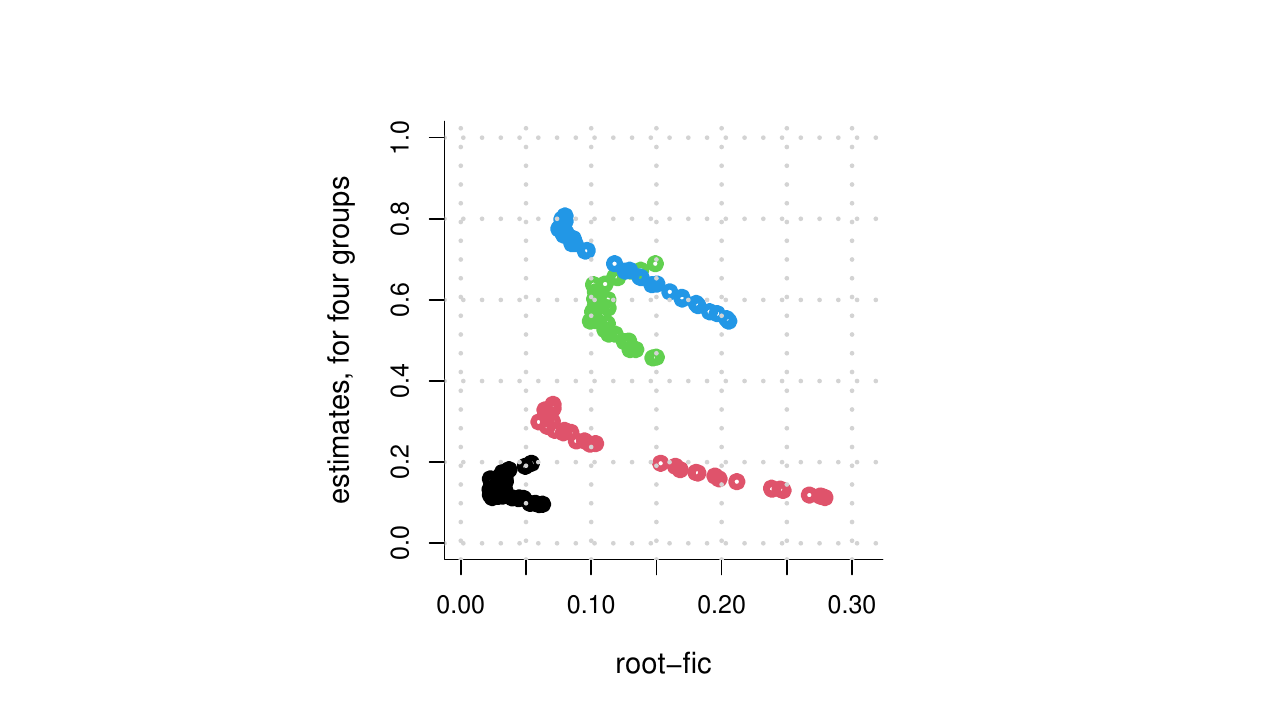} 
  \caption{FIC plots for four strata of individuals,
stratum 1 (black), 2 (red), 3 (green) and 4 (blue).
The estimates, associated with the $2^5=32$
candidate models, are on the vertical axis, with root-fic on the
horizontal, i.e.~estimated root-mse. The more to the left
in the plot, the better the submodel and its estimate.
There are different best models for the different strata.}
\label{figure:eyes}
\end{figure}

The AIC solution is found to include $z_3,z_5$, hence ignore
$z_1,z_2,z_4$, for all estimation purposes;
the FIC spirit is however to identify different best models
for different problems.
Motivated by attempting to understand
which factors are important for assessing the risk of
diabetic retinopathy, for those with relatively high values
of bmi and pulse rate, we pick individuals with
median value for $x_1,z_1$ but at 90 percent quantile levels
for $z_2,z_3$, with $z_4=1$ (i.e.~male), and then
the four possibilities for $x_2$ and $z_5$
(presence or not of edema; presence or not of urine protein).
Below we refer to these four strata as stratum 1 with $(0,0)$,
stratum 2 for (0,1), stratum 3 for (1,0) and stratum 4 for $(1,1)$.
Details concerning how to actually compute the FIC values
using the R library \texttt{fic} appear
in Section~\ref{subsection:ficA-compute}.

\textbf{Figure \ref{figure:eyes}} is an instance of so-called {\it FIC plots}.
There are variations here, but this follows a certain standard,
with candidate model estimates on the y-axis
and root-FIC scores on the x-axis; the more to the left,
the better the estimate, and vice versa.
Also, the root-FIC has direct interpretation as
estimating the root-mse.
Going through the different cases we learn the following.
For stratum 1, the best is to add $z_2$ with focus estimate 0.1332;
for strata 2 and 4, the FIC selects $z_3,z_5$, i.e.~the AIC winner. The corresponding focus estimates are 0.2991 for stratum 2 and 0.7751 for stratum 4; for stratum 3, adding $z_1$ is the FIC preference, with focus estimate 0.5476.

\subsection{Computing the FIC value}
\label{subsection:ficA-compute}

The R package fic \citep{Jackson25} computes the value of the FIC for focused variable selection questions concerning generalized linear models, the Cox proportional hazard regression model, (flexible) parametric survival models and multistate models for panel data.
The required input for the function \texttt{fic()} in this R package is:
\begin{enumerate}
\item The fitted model output for the largest model (e.g.~the result of using \texttt{glm()}).
\item The set of models to search from. This can be user specified, a nested sequence of models, or all subsets of the largest model.
\item Stating which parameters are protected (this is the \texttt{inds0} value).
\item Defining the focus as a function of the model parameters. Some common choices are included in the \texttt{fic()} function, or this can be user specified.
\item The covariate value(s) at which the focus is to be computed.
\end{enumerate}
In this R package the wide model estimates are used to estimate the parameter values that occur in the final FIC formula.

For the Diabetic retinopathy study in Section~\ref{subsection:klein2008},
FIC selection is performed as follows.
\begin{small}
\begin{verbatim}
library(fic)
wide = glm(y ~ x1+x2+z1+z2+z3+z4+z5, family=binomial)
inds0 = c(rep(1,3),rep(0,5))  # first 3 parameters protected, other 5 open
X.eval = rbind("stratum1"=c(1,10,0,10.6,28,50,1,0),
        "stratum2"=c(1,10,0,10.6,28,50,1,1),
        "stratum3"=c(1,10,1,10.6,28,50,1,0),
        "stratum4"=c(1,10,1,10.6,28,50,1,1))
combs = all_inds(wide,inds0)  # all subsets search
fic1 = fic(wide,inds=combs,inds0=inds0,focus="prob_logistic",X=X.eval)
summary(fic1,adj=TRUE)
\end{verbatim}
\end{small} 
This produces the following summarized output. More details about the fitted models, the estimated bias, variance, and focus values for each of the models are computed too and accessible to the user, but these are not shown in the summary.
Note that the value of the average FIC, AFIC, see Section~\ref{subsection:ficA-AFIC}, is included in the summary when more than one evaluation value is specified.
\begin{small}
\begin{verbatim}
Model with lowest RMSE by focus
         index                    pars     focus
stratum1     3    (Intercept),x1,x2,z2     0.1332
stratum2    21 (Intercept),x1,x2,z3,z5     0.2991
stratum3     2    (Intercept),x1,x2,z1     0.5476
stratum4    21 (Intercept),x1,x2,z3,z5     0.7751
Average     21 (Intercept),x1,x2,z3,z5     0.4384

Range of focus estimates and RMSE over models
         min(focus) max(focus)  min(RMSE)  max(RMSE)
stratum1 0.0954     0.1972      0.0225     0.0627
stratum2 0.1127     0.3424      0.0599     0.2792
stratum3 0.4570     0.6891      0.0993     0.1499
stratum4 0.5477     0.8069      0.0751     0.2054
Average  0.3302     0.4769      0.0718     0.1802
\end{verbatim}
\end{small}

Several worked out data examples are included in the R package's vignette files. These include focused selection among skewed normal models, survival models and generalized linear models. Examples of focus functions are given, which can be completely user-specified, and it is explained how one can specify a loss function different from the least squares loss.
For a graphical representation of the output, see the FIC plots in
Section~\ref{subsection:klein2008}.

For focused selection in other model classes not contained in the library, several authors have developed their own functions. Each time, the main ingredients are the same as above: the Fisher information matrix in the largest model, the estimator of $\gamma$ in the largest model and the partial derivatives of the focus function with respect to the model parameters.

\subsection{FIC extensions}
\label{subsection:ficextenstions}

First, the covariates in a regression model can enter
the model in several ways. To illustrate
this point, consider a broader version
of the logistic regressions exhibited in
Section \ref{subsection:klein2008}, 
\beqn
p(\bm{x},\bm{z},\bm{w})
=H(\bm{x}^\tr\bbeta+\bm{z}^\tr\bgamma)^{\exp(\bm{w}^\tr\bkappa)},
\eeqn
allowing skewing the logistic transform, via
a subset $\bm{w}$ of the covariates $(\bm{x},\bm{z})$.
The previous setup corresponds to $\bkappa$
being a vector of zeros only.
With $r$ the length of $\bm{w}$,
the general FIC theory still applies, with the possibility
of searching through $2^{q+r}$ models and selecting variables $z_j$ and $w_j$ for inclusion
or exclusion in respectively $\bm{z}^\tr\bgamma$
and $\bm{w}^\tr\bkappa$.
This necessitates new formulae 
for $\bm{J}_\wide$
and $\bomega$ of equations (\ref{eq:Jwide}) and (\ref{eq:tau0andomega}),
but otherwise follows the general algorithm.
Carrying out extended FIC analysis for the
setup of Section \ref{subsection:klein2008}, but
now allowing $\bm{w}^\tr\bkappa$ above as skewness
modificators, using e.g.~$\bm{w} = (x_1,x_2,x_3)^\tr$
with three more parameters, would lead to a more crowded FIC plot
than in \textbf{Figure \ref{figure:eyes}}, now with $2^{5+3}=256$
point estimates for each stratum. Again, FIC finds
the best candidate model for each.

Other situations of this type are heteroscedastic
normal regression, with
$Y_i\sim\N(\bm{x}_i^\tr\bbeta,\sigma^2\exp(\bm{z}_i^\tr\bgamma))$,
and doubly log-linear gamma regression, with
$Y_i\sim\Gam(a_i,b_i)$, say, where $a_i=\exp(\bm{x}_i^\tr\bbeta)$
and $b_i=\exp(\bm{x}_i^\tr\bgamma)$.

Second, the FIC naturally works also for i.i.d.~setups,
for two-sample comparisons, etc. For illustration,
suppose there are data $x_i$ from $F$ and $y_i$ from $G$,
with wide models $(\btheta,\bgamma)$ of dimension $p+q$,
narrow models $(\btheta,\bgamma_0)$, for each,
with $\bgamma_0$ a fixed value of $\bgamma$, like zero.
To estimate the median difference
$\mu=\med(G)-\med(F)$, for example, one may use the
natural variations and extensions of eq.~(\ref{eq:master1})
to build a FIC, for this focus parameter,
to select the best among $2^{q+q}$ candidate models.

\section{NARROW VS.~WIDE, TOLERANCE RADII, AFIC}
\label{section:aficandmore}

The FIC schemes above were constructed from the basic
Master Theorem in eq.~(\ref{eq:master1}). Other relevant insights,
of separate interest, also flow from that and related results.

\subsection{Tolerance around models}
One question to ask is whether extending a narrow model is needed.
For a given dataset, should we be content with a perhaps
classical regression model, with $p$ parameters, or does it pay off
to include one more, to signal model departure in a certain
direction, or even $q$ more? From the FIC perspective,
this depends not on `overall fit' to data, but on
the specific question for which we need the best answer.
It is indeed useful to check the special cases of the narrow and
wide model in the FIC framework.
These correspond to $S=\emptyset$, with $G_S=0$,
and $S=\{1,\ldots,q\}$, with $\bm{G}_S=\bm{I}_q$, and we have
$
\mse_\narr=\tau_0^2+(\bomega^\tr\bdelta)^2
$
and 
$
\mse_\wide=\tau_0^2+\bomega^\tr \bm{Q}\bomega,
$ 
clearly illustrating the bias-variance balance.
The narrow model is best when
$|\bomega^\tr\bdelta|\le(\bomega^\tr \bm{Q}\bomega)^{1/2}$,
which is an infinite tolerance strip around the narrow model,
in terms of $\bdelta=\rootn(\bgamma-\bgamma_0)$.
For some focus parameters, it would not be alarming that
the narrow model could be fully correct; for others,
it would matter significantly.

For the one-dimensional case, the $\omega$ factor cancels out,
and the above simplifies to the tolerance radius
$|\gamma-\gamma_0|\le\kappa/\rootn$,
with $Q=\kappa^2$. It is instructive to go through
various settings, with a perhaps familiar model
$f(y;\btheta)$ as the narrow model, and then investigating
whether it pays off to include one more parameter,
say with $f(y;\btheta,\gamma)$ as the wide model.
For such situations it is then a matter of computing
the $(p+1)\times(p+1)$ Fisher information matrx $\bm{J}_\wide$,
and then read off the $Q=J^{11}=\kappa^2$.
If the tolerance radius $\kappa/\rootn$ is big,
in the relevant context, it might not be worth the
extra modelling effort, and vice versa. We also learn
in a precise fashion the role of increasing sample size;
for large $n$, it pays off to use more sophisticated
models. Various examples, from i.i.d.~setups to regression
models, are given in \citet[Ch.~3]{ClaeskensHjort08}
and \citet{Hjort93}.

Going back to the FIC selection, for the limit experiment
worked with above, we have
\beqn
\fic_\narr=\tau_0^2+\bomega^\tr(\bm{DD}^\tr-\bm{Q})\bomega
\quadandquad
\fic_\wide=\tau_0^2+\bomega^\tr \bm{Q}\bomega.
\eeqn
So FIC prefers the narrow to the wide provided
$(\bomega^\tr \bm{D})^2 \le 2\bomega^\tr \bm{Q}\bomega$.
We again see that different choices are made for different foci,
i.e.~different $\bomega$.
As for the tolerance radius discussion,
matters simplify for the one-dimensional case $q=1$,
with the $\omega$ cancelling out;
here the narrow model is preferred as long as $|D|\le 2^{1/2}\kappa$,
writing as above $Q=\kappa^2$. For the real-data setup,
this translates to $|\hatt\gamma-\gamma_0|\le 2^{1/2}\kappa/\rootn$.
In the large-sample sense, this is seen to be equivalent
to the AIC. So `FIC = AIC', essentially, for the case of
comparing a start model with a one-parameter extension,
but not for $q\ge2$.

\subsection{AFIC, the average-FIC}
\label{subsection:ficA-AFIC}

Often there is not a single focus parameter,
but a collection of such, e.g., all quantiles above level 0.90,
the mean response 
for all individuals
inside a certain stratum.
In general terms, suppose there are foci $\mu_j$ to be estimated,
along with weights of importance $v_j$, for $j=1,\ldots,k$.
With candidate model $S$, estimators $\hatt\mu_{S,j}$
are computed, with overall loss
$L_S=\sum_{j=1}^k v_j(\hatt\mu_{S,j}-\mu_{j,\true})^2$,
i.e.~the weighted sum of squared errors. For each focus,
there is an $\bomega_j$, from the expressions
in eq.~(\ref{eq:tau0andomega}). Using Master Theorem I of (\ref{eq:master1}),
there is a clear limit
$
nL_S\arr_d \Lambda_S=\sum_{j=1}^k v_j\{\Lambda_{0,j}
   +\bomega_j^\tr(\bdelta-\bG_S \bD)\}^2,
$
with limiting risk its mean
\beqn
r_S=r_0 + \sum_{j=1}^k v_j [\bomega_j^\tr {\bG}_S \bQ {\bG}_S^\tr\bomega_j
+\{ \bomega_j^\tr(\bm{I}_q-{\bG}_S)\bdelta\}^2 ]
=r_0+ \Tr [\{{\bA}_S+{\bB}_S(\bdelta)\} \bC ],
\eeqn
writing $r_0=\sum_{j=1}^k v_j \tau_{0,j}^2$ for the term
not depending on $S$, along with
${\bA}_S={\bG}_S \bQ {\bG}_S^\tr$,
${\bB}_S(\bdelta)=(\bm{I}_q-{\bG}_S)\bdelta\bdelta^\tr(\bm{I}_q-{\bG}_S)^\tr$,
and $\bC=\sum_j v_j\bomega_j\bomega_j^\tr$.
For model comparison purposes we may ignore the $r_0$ term.
Our AFIC, the weighted or average FIC, given the list
of foci and their weights of importance, is then, with ${\bM}_S$ estimating $\bB_S(\bdelta)$,
\beq
\AFIC_S=\Tr({\bA}_S {\bC}) + \Tr({\bM}_S \bC).
\label{eq:aficA}
\eeq For this limiting experiment,
the unbiased estimator is
$
{\bM}_S=(\bm{I}_q-{\bG}_S)(\bD \bD^\tr-\bQ)(\bm{I}_q-{\bG}_S)^\tr
   =\{(\bm{I}_q-{\bG}_S)\bD\}^2 - (\bm{I}_q-{\bG}_S) \bQ (\bm{I}_q-{\bG}_S)^\tr,
$
which we choose truncate to zero in case its value is negative.

Just as for the single-FIC discussed in
Section~\ref{subsection:ficA-deffocus}, a `real AFIC' emerges
by plugging in estimators for unknown quantities,
involving the observed Fisher information matrix
$\hatt{\bJ}_\wide$, and using ${\bD}_n=\rootn(\hatt{\bgamma}-{\bgamma}_0)$
for $\bm{D}$. The theory works also when the weights $v_j$
are not given a priori, but rather estimated consistently
from data, as with $v_j=v_j(\btheta,\bgamma)$
for an appropriate smooth $v_j$ function.
See \citet[][Remark 6.4]{ClaeskensHjort08}
for a connection between AFIC and AIC.-

It is useful to spell out how the narrow and wide models
compare here. The risks simplify to
$r_\narr=\bdelta^\tr \bC \bdelta$, $r_\wide = \Tr(\bQ \bC)$,
ignoring the first term $r_0$ that does not depend on the $S$,
and
$
\AFIC_\narr=\Tr\{ \bC (\bD \bD^\tr-\bQ) \} = \bD^\tr \bC \bD-\Tr(\bC \bQ),
$
$
\AFIC_\wide = \Tr(\bC \bQ).
$
In the limit experiment, narrow is judged best
when $\bD^\tr \bC \bD \le 2\,\Tr(\bC \bQ)$.
In some settings the foci $\mu_j$ and their weights of importance
$v_j$ are arranged such that $\bC = \bQ^{-1}$,
see eq.~(\ref{eq:hereisQ}), in which case matters simplify further;
narrow is then preferred as long as $\bD^\tr \bQ^{-1} \bD \le 2q$.
This precisely matches the classic AIC,
see \citet[Ch.~2]{ClaeskensHjort08}. Briefly, with
$
\AIC_S=-2\ell_{n,S}(\hatt{\btheta}_S,\hatt{\bgamma}_S,{\bgamma}_{0,S^c})+2|S|,
$
there are precise limits for $\AIC_\narr-\AIC_S$,
in terms of quadratic functions of $\bD$ and additional terms,
with consequent decision boundaries and model ranking.
These are identical to those for the AFIC, if and only
if $\bC=\bQ^{-1}$.
For the important class of generalized linear models (GLM),
details related to these issues are given in
\citet{ClaeskensHjort08b}, also pointing to which
weights, depending on the distribution, achieve
large-sample equality with the AIC. For linear regression
models, with means
$\E\,(Y_i\midd \bx_i,\bz_i)=\bx_i^\tr\bbeta + \bz_i^\tr\bgamma$,
the $\bx_i$ protected and $\bz_i$ open,
this amounts to giving the same weight $1/n$
to each focus $\bx_i^\tr\bbeta + \bz_i^\tr\bgamma$.
For various other cases, the AFIC has a real focus,
different from the bland overall goal of the AIC;
AFIC can be seen as properly focused
versions of the AIC.
The FIC is more versatile than the AIC, however, by allowing
user-specified weights (including equal weighting also
for non-normal distributions) and at the same time by directing
the model search towards the focus that is to be estimated accurately.

The default option in the R package \texttt{fic} is to use equal weights when more than one focus is specified, though user-specified weights can be passed on to the function for the computation of the AFIC.
For the diabetic retinopathy data of Section~\ref{subsection:klein2008}, we specify as focuses the $p(\bm{x},\bm{z})$ for all 25 subjects in stratum 3 with $x_2=1$ and $z_5=0$. Equal weights with the value 1/25 are used. The information for the best 3 models is shown below. The column \texttt{mods} indicates which variables are present in the model, in the order as specified in the wide model, the intercept is included.  The root mean squared error, bias and standard error values are formed as per the formulas for AFIC, including the bias correction and truncation to zero for the squared bias part. The best model according to AFIC for the subjects in stratum 3 is the model with intercept, $x_1$, $x_2$, $z_2$, $z_3$ and $z_5$. The AIC selected model comes at a close second place with nearly identical root mean squared error.
\begin{small} 
\begin{verbatim}
   vals     mods rmse.adj    bias     se  focus
Average 11101101   0.0858 -0.0117 0.0850 0.6779
Average 11100101   0.0859  0.0196 0.0836 0.6841
Average 11110001   0.0861  0.0131 0.0851 0.6912
\end{verbatim}
\end{small}

\section{FIC AND AFIC WITH A FIXED WIDE MODEL}
\label{section:ficB}

In Section~\ref{section:ficA} a broadly
applicable FIC tool is described
for i.i.d.~and regression models.
The operating condition is that the competing models
are not very different, with parameters scaled to the $O(1/\rootn)$
setup, corresponding to 
$\|\rootn(\bgamma-\bgamma_0)\|$
inside a bounded terrain. Here we study a different setup,
most easily presented in the i.i.d.~situation where
$Y_1,\ldots,Y_n$ stem from some unknown distribution
with density $f$ and c.d.f.~$F$. When estimating
a focus quantity $\mu=\mu(F)$, like the median, skewness,
the probability $F(b)-F(a)$, etc., there is a choice
between the wide model nonparametric  $\hatt{\mu}_\nonpara=\mu(F_n)$,
writing $F_n$ for the empirical distribution with mass $1/n$
on each datapoint, and parametric
$\hatt{\mu}_\para=\mu(F(\cdot,\hatt{\btheta}))$,
using ML estimation for different competing candidate models.

\subsection{mse approximations and FIC formulae}
\label{subsection:jullumhjort}

The FIC approach here, for comparing parametric with nonparametric estimators,
is to follow the main steps of eq.~(\ref{eq:msebasic})
and (\ref{eq:ficbasic}), but with formulae different from
those obtained in Section~\ref{section:ficA}.
The essence is as follows. Under regularity conditions,
estimators $\hatt\mu_\nonpara$ and $\hatt\mu_\para$
will have a binormal approximation. A formalized version
is that
\beq
\begin{array}{rcl}
\hatt\mu_\para
&=&\displaystyle
   \mu(\btheta_0)+c_\para/n+n^{-1/2} Z_{n,\para}+o(1/n), \\
\hatt\mu_\nonpara
&=&\displaystyle
   \mu_\true+c_\nonpara/n+n^{-1/2} Z_{n,\nonpara}+o(1/n),
\end{array}
\label{eq:jullumhjortA}
\eeq
with $(Z_{n,\para},Z_{n,\nonpara})$ having a joint zero-mean
binormal limit, with variances $v_\para$ and $v_\nonpara$,
and covariance $k$.
Here $\btheta_0$ is the least false parameter the ML estimator
$\hatt{\btheta}$ is aiming for, the minimizer of the Kullback--Leibler
distance $\KL(f,f_{\scriptsize\btheta})$. This leads to approximations
\beq
\mse_\para=b^2+2bc_\para/n+v_\para/n, \quad
\mse_\nonpara=0^2+v_\nonpara/n,
\label{eq:jullumhjortB}
\eeq
with $b=\mu(\btheta_0)-\mu_\true$.
Note that with
$\hatt b=\hatt\mu_\para-\hatt\mu_\nonpara$, we have
$\rootn(\hatt b-b)$ equal to $Z_{n,\para}-Z_{n,\nonpara}$ plus
smaller terms, tending to a zero-mean normal with
variance $w=v_\para+v_\nonpara-2k$.
These assessments lead to FIC scores, by estimating
the required quantities,
\beq
\fic_\nonpara=\hatt v_\nonpara/n,
\quad
\fic_\para=\max(\hatt b^2+2\hatt b\hatt c_\nonpara/n-\hatt w/n,0)
   +\hatt v_\para/n.
\label{eq:ficficwide}
\eeq
Importantly, these risk estimators need to work
well in the widest model, here the nonparametric setting,
i.e.~outside the conditions of the parametric models.
For various cases it is convenient to present these
formulae in terms of the influence functions
for the two types of estimators;
see Supplement sections A, B 
for details. For further discussion and applications,
see \citet{JullumHjort17}, \citet{DaehlenHjortHobaekhaff24}.

\subsection{Illustration:
  Sample quantiles vs.~parametrics for birthweights distribution}
\label{subsection:oslokids}

\begin{figure}
\begin{subfigure}[t]{0.48\linewidth}
\includegraphics[width=\linewidth]{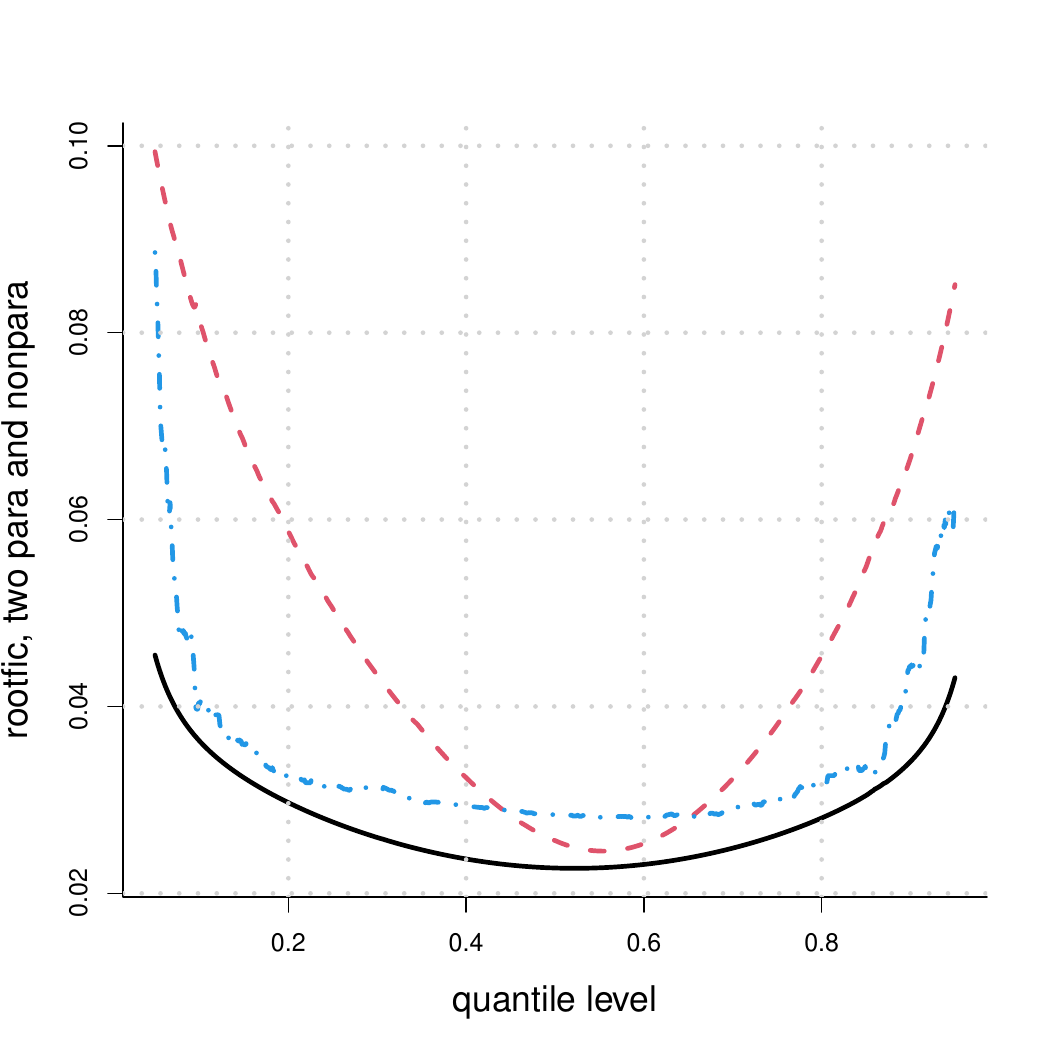}
\caption{Root-FIC scores, as functions of quantile level
    $q\in[0.05,0.95]$, for estimating $F^{-1}(q)$,
    for the direct sample quantile (wiggly),
    and for the parametric alternatives normal (dashed)
    and t (full curve). The t model is best.}
\label{figure:babiesA}
\end{subfigure}
\hfill
\begin{subfigure}[t]{0.48\linewidth}
\includegraphics[width=\linewidth]{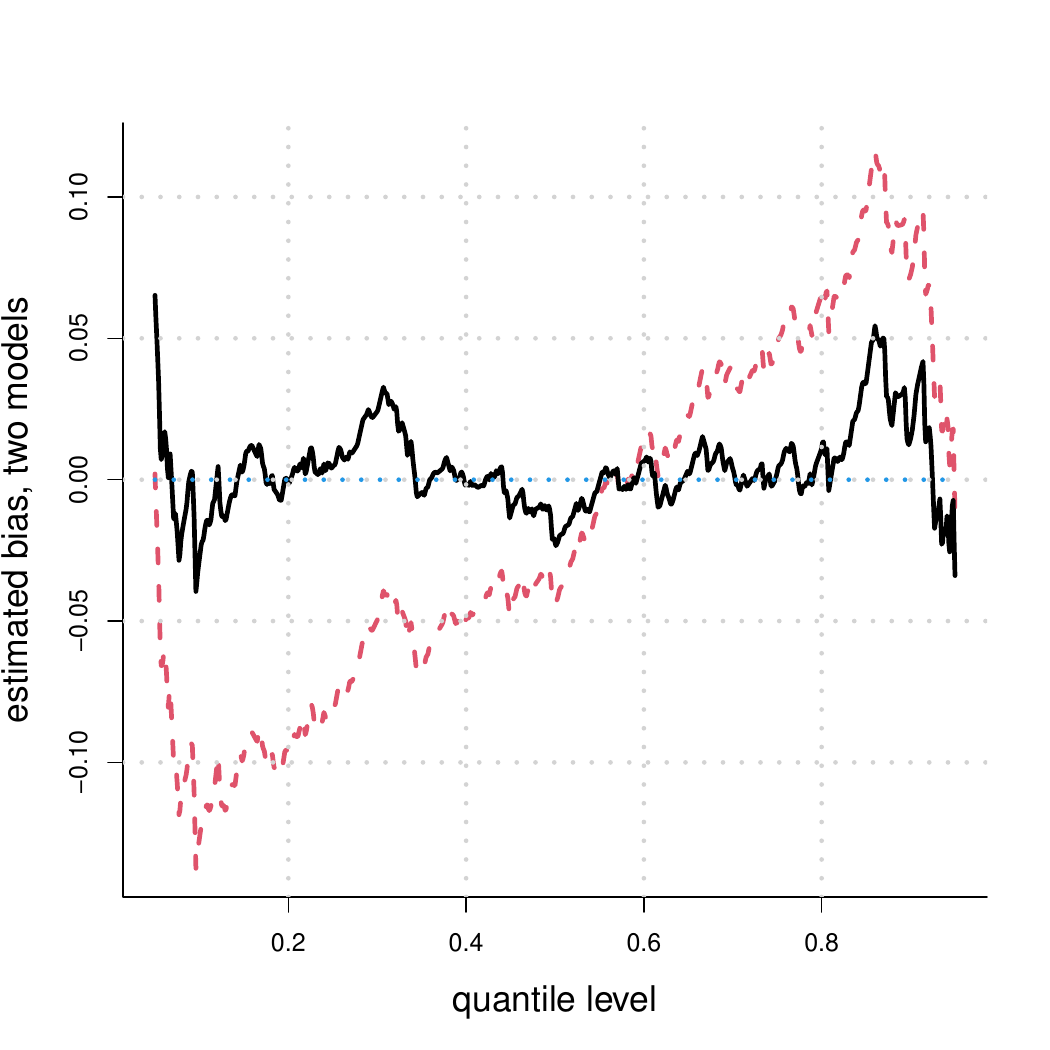}
\caption{
  Estimated biases, for the normal (dashed, sizeable)
  and t based (full curve, close to zero) quantiles.}
\label{figure:babiesB}
\end{subfigure}
\caption{Root-FIC analysis for the birthweight quantiles
  for Oslo born girls, with two parametric models
  competing with the nonparametric one.}
\label{figure:babies}
\end{figure}

For $n=480$ girl birthweights $y_1,\ldots,y_n$,
recorded at Rikshospitalet in Oslo 2001--2008, we wish to estimate
the quantile $\mu_q=F^{-1}(q)$ of the underlying distribution.
We consider three different estimators:
the direct nonparametric sample quantiles, and
the parametric alternatives using the normal $\N(\xi_0,\sigma_0^2)$
and the three-parameter t distribution, say $t(\xi,\sigma,\nu)$.
This leads to
$
\hatt\mu_{q,\nonpara}=Q_n(q),\quad
\hatt\mu_{q,N}=\hatt\xi_0+\hatt\sigma_0\,\Phi^{-1}(q),\quad
\hatt\mu_{q,t}=\hatt\xi+\hatt\sigma\,G^{-1}(q,\hatt\mu),
$
with parameters estimated via ML; 
one finds $(3.494,0.562)$ for the normal
and $(4.515,0.428,4.586)$ for the t.
The sample quantile $Q_n(q)$ is approximately unbiased,
with variance $\kappa_q^2/n=y(1-y)/\{nf(\mu_q)^2\}$,
in terms of the density $f$.
The FIC task is to estimate the
mean squared errors $\mse(q)$ for the three estimators.
The root-FIC scores are displayed in \textbf{Figure~\ref{figure:babiesA}}.
It is informative here to check the modelling biases,
displayed in \textbf{Figure~\ref{figure:babiesB}};
we learn that quantile estimators using the normal model
have a sizeable bias, whereas the t model has small bias.
For the sample of birthweights of the Oslo girls
the t distribution is the best choice.
See Supplement Section C 
for some of the details pertaining to estimation of the $\kappa_q$,
and \citet[Story \#23]{HjortStoltenberg26} for yet further
related themes.

Similar to 
Section~\ref{subsection:ficA-AFIC}, we can build
AFIC scores also in the present fixed wide model framework,
to choose the best model for an ensemble of
focus parameters $\mu_j$, with context based
weights of importance $v_j$ given to these.
We might e.g.~wish a good model for birthweights
for a range of upper quantiles, with increasing $v_j$
for quantiles $q_j$, and perhaps with
further parametric candidate models; AFIC will then
do such a job.

\subsection{CAT-FIC: for categorical data}
\label{subsection:catfic}

FIC schemes can be worked out for contingency tables,
multinomial data, and other categorical data settings.
Suppose $(N_1,\ldots,N_k)$ is a multinomial $(n,p_1,\ldots,p_k)$,
with parameter $\mu=\mu(p_1,\ldots,p_k)$
to focus on. The nonparametric
$\hatt\mu_\nonpara=\mu(\tilda p_1,\ldots,\tilda p_k)$,
with direct estimators $\tilda p_j=N_j/n$, is approximately
unbiased, normal, and with variance $v_\nonpara/n$, with
$v_\nonpara=\sumjk c_j^2p_j-(\sumjk c_jp_j)^2$.
A parametric model, say $p_j=p_j(\btheta)$, can lead to the
better estimator $\hatt\mu_\para=\mu(\hatt p_1,\ldots,\hatt p_k)$,
with $\hatt p_j=p_j(\hatt{\btheta})$, with the ML estimator,
by having a smaller variance and perhaps not a big bias.
The latter is approximately $b=\mu({\btheta}_0)-\mu_\true$,
writing $\mu(\btheta)=\mu(p_1(\btheta),\ldots,p_k(\btheta))$,
and with $\btheta_0$ the least false parameter value.

This becomes a job for FIC, setting up estimators
for $v_\nonpara/n$ and the appropriate $v_\para/n+b^2$;
details and examples are given in \citet[Section 6]{JullumHjort17}.
This makes it possible to check, e.g., whether
parameters 
are better estimated under independence assumptions,
than without (even when strict independence
does not hold). There are similarly AFIC versions
for contingency tables and multinomial data.
Suppose the precision of the $\hatt p_j=p_j(\hatt{\btheta})$
are to be compared with the raw data frequencies $\tilda p_j$,
using the combined loss $L=n\sumjk (p_j^*-p_{j,\true})^2/p_{j,\true}$.
There are associated risks $r_\para$ and $r_\nonpara$,
the latter being as easy as $k-1$. The usual Pearson
goodness-of-fit statistic in such a setup is
$
X_n=n\sumjk{ (\hatt p_j-\tilda p_j)^2/ \tilda p_j},
$
tending under parametric model conditions to a $\chi^2_\df$,
with $\df=k-1-q$, with $q$ the dimension of $\theta$.
Here we need approximations to $X_n$ and its mean
under wider nonparametric conditions. Some work leads to
\beqn
\afic_\para=X_n-(k-1)+2q^*,
\quad {\rm with\ }
q^*=\Tr(\hatt J^{-1}\hatt J^*),
\eeqn
with expressions for the  $q\times q$ matrices
$\hatt J$ and $\hatt J^*$ given in
Supplement Section D. 
Under model conditions, these two matrices aim for the
same underlying $J$ matrix, and $q^*\arr_\pr q$.

So when is a parametric model under consideration good enough,
i.e.~better than the wide nonparametric model?
The parametric is better, in terms of the AFIC,
provided $X_n-(k-1)+2q^*\le k-1$, i.e.~$X_n\le 2(k-1-q^*)$.
We learn that the FIC and AFIC viewpoint, assessing
models via their risks, amounts to an implied
goodness-of-fit test, though without any consideration
of null distribution of the $X_n$, and without any
extraneous 0.05 significance level. As such the $X_n$
based test could be seen as a \textit{clearance test}, without
the usual extra language of null hypotheses and p-values etc.
Incidentally, if the model actually holds, then
for the rejection probability we have
$\Pr_{\rm model}({\rm rejected})\arr\Pr(\chi^2_\df \ge 2\cdot\df)$,
which in testing language means the significance level.
The first few of these values are 0.157, 0.135, 0.112, 0.092, 0.075,
for $\df$ from 1 to 5.

\subsection{General FIC with fixed wide model}

In Section \ref{subsection:jullumhjort} we presented
the FIC for parametric vs.~nonparametric estimators for i.i.d.~set\-ups,
illustrated with the Oslo girls in Section \ref{subsection:oslokids},
there comparing two parametric models against a nonparametric model
for estimating quantiles. Similar constructions
may be put up also for regression models and indeed for
more complex setups, as we explain here.

For illustration, consider regression type data $(x_i,Y_i)$,
following the signal plus noise structure
$Y_i=m(x_i)+\eps_i$, with $\N(0,\sigma^2)$ errors.
Here we may be interested in comparing $m_1(x)=a+bx$,
$m_2(x)=a+bx+cx^2$,
against the nonparametric alternative where $m(x)$
is simply taken to be an unknown smooth function.
Focus parameters could include $\mu=m(x_0)$ at a given
location $x_0$, or the probability $\Pr(Y_0>y_0\midd x=x_0)$, etc.
This is tricky terrain, (i) since one needs to define
a nonparametric smoother, say $\hatt m(x)$, perhaps
using kernel methods, with tuning parameters to decide on;
and (ii) since it becomes more difficult
both to approximate biases and variances
and to estimate these. This is doable, as indicated
in \citet[Section 7]{JullumHjort17}, but becomes intricate,
also encountering different factors $1/n$ and $1/n^{4/5}$
for the variances.

It is relatively speaking easier, for many such problems,
particularly for more complicated parametric competing models,
to avoid the too broad `fully nonparametric' wide model,
but to relate competitors to a carefully constructed
`fixed parametric wide model'. For the signal plus noise
illustration above, such a fixed model could be the
cubic regression model with $m_3(x)=a+bx+cx^2+dx^3$.
It is then possible to assess and estimate the biases
and variances of the two models $m_1(x)$ and $m_2(x)$,
leading, once more, via basic steps (\ref{eq:msebasic})
and (\ref{eq:ficbasic}) to FIC procedures, for different
types of focus parameters.

In linear mixed models of the form $\bm{Y}_{i} = \bm{x}_i\bbeta+\bm{z}_i\bm{U}_i + \bvarepsilon$ where the response groups all $m_i$ observations for observation $i$ (with $i=1,\ldots,n$) next to the fixed effects parameter $\bbeta$ there is a vector of random effects $\bm{U}_i$, independent of the error terms $\bvarepsilon_i$. Variable selection questions may involve both $\bbeta$ and the often normally distributed $\bm{U}_i$ with mean zero. Absence of a random effect occurs when its variance equals zero, which is a value at the boundary of its domain. For this reason, \cite{CunenWalloeHjort20} used a fixed wide model to develop the FIC for linear mixed models. See also the application in \citet{CunenWalloeKonishiHjort21}.

\section{USING FIC TO DETERMINE FINE-TUNING PARAMETERS}
\label{section:fic-finetuning}

Above our focus has been that of using FIC to
select among candidate models. 
What is being
compared and ranked are not the models $M_j$,
per se, but their outputs, the estimates $\hatt\mu_j$
of a common quantity $\mu$. The FIC idea hence also applies
to other statistical selection tasks. An important
category of such situations is when constructing
estimators $\hatt\mu_a$ for a focus $\mu$, with $a$
a tuning parameter. In generic terms, there is a
$\mse_a=\E_\true[(\hatt\mu_a - \mu_\true)^2]$,
just as with eq.~(\ref{eq:msebasic}), and if we manage
to construct estimators $\fic_a=\hatt{\mse}_a$,
as with eq.~(\ref{eq:ficbasic}), we can select $a$
as the minimizer of this $\fic_a$.

\begin{figure}
\begin{subfigure}[t]{0.48\linewidth}
\includegraphics[width=\linewidth]{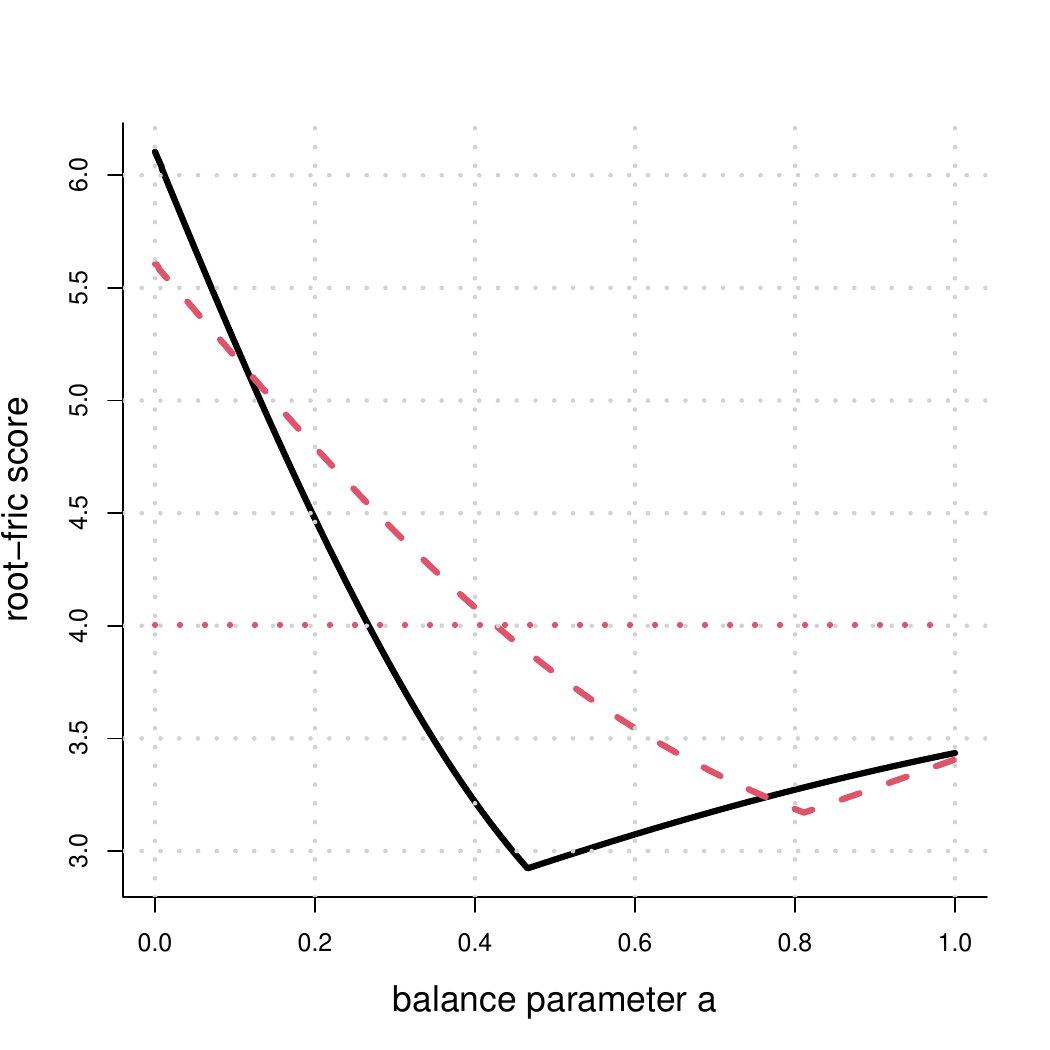}
\caption{The $\fric(a)^{1/2}$ curves, for the gamma model
  (smooth, black) and Weibull model (slanted, red),
  with the horizontal line at $4.07$ being the root-FIC
  for the nonparametric estimator. The ML estimators
  are worse, but the BHHJ estimators are better,
  with optimal balance parameters 0.464 and 0.807
  for the two models.}
\label{figure:egyptA}
\end{subfigure}
\hfill
\begin{subfigure}[t]{0.48\linewidth}
\includegraphics[width=\linewidth]{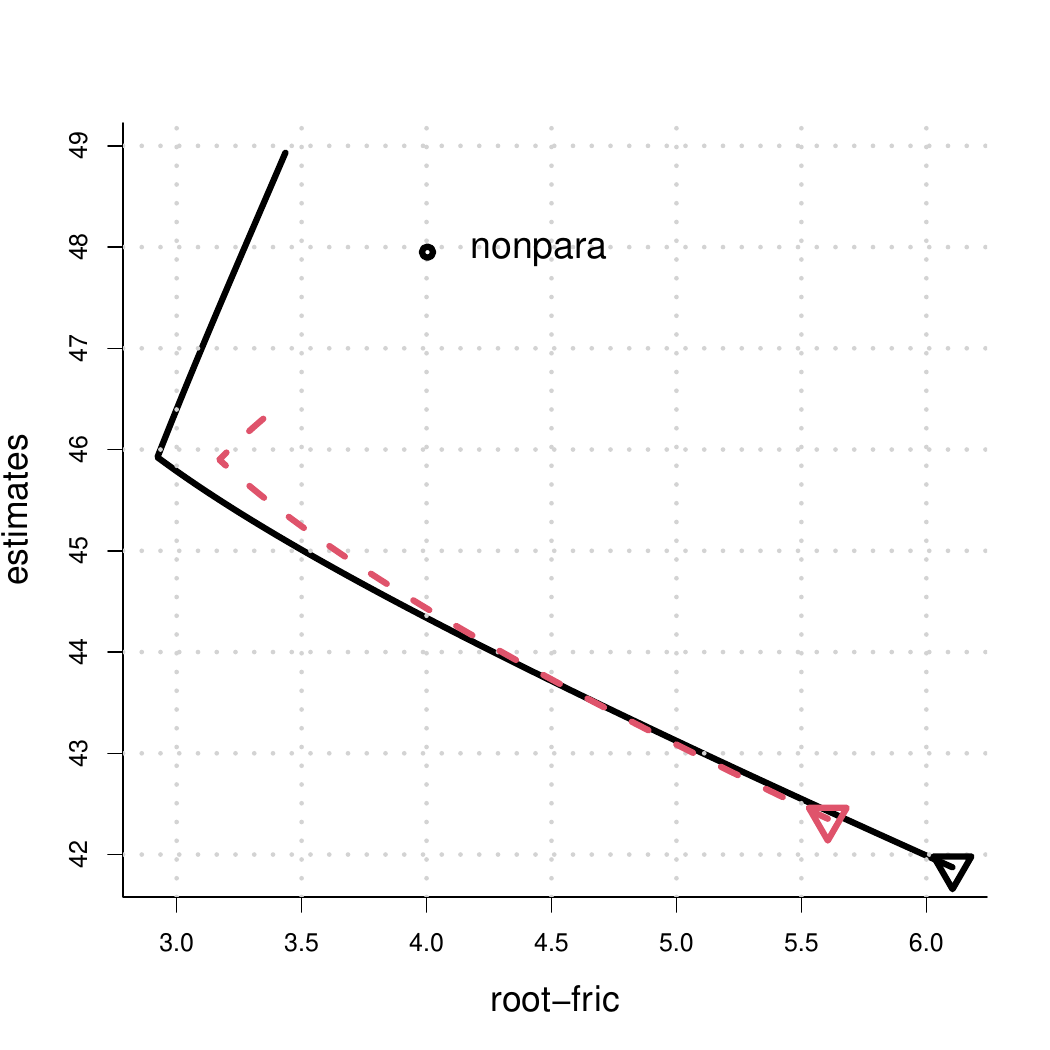}
\caption{The estimates $\hatt\mu_\para$ on the
  y axis, plotted against the root-FIC scores on the x axis,
  along with the nonparametric $\hatt\mu_\nonpara=48.0$.
  The triangles indicate where the BHHJ machine
  starts, with the ML estimates at $a=0$.
  The best estimates are almost equal
  for the gamma and the Weibull, $45.91$ and $45.88$,
  with the first being the very best,
  according to the FRIC analysis.}
\label{figure:egyptB}
\end{subfigure}
\caption{Analysis for estimating the 0.75 quantile
  of the lifelength distribution in Roman Era Egypt,
  with the $a$ BHHJ parameter taken to range from 0 to 1.
  Both the gamma and the Weibull models lose to
  the nonparametric when ML estimators are used,
  i.e.~$a=0$, but with BHHJ estimators both parametric
  models do better than the nonparametric $\hatt\mu_\nonpara=48.0$.}
\label{figure:egypt}
\end{figure}

Assessing or approximating the $\mse_a$, followed by
construction of the $\fic_a$, depends on the framework
and the $\hatt\mu_a$ estimator, but also on what
is taken to be the underlying data generating process.
An important setup where these ideas can be implemented
is that of Section \ref{section:ficB}. For the FIC
developed there we relied on ML estimators to form $\hatt\mu_\para=\mu(\hatt{\btheta})$
for the different candidate models, of type $f(y,\theta)$.
These are optimal under the conditions of the model worked with,
but are both prone to non-robustness and to model
misspecification. To show how the FIC schemes
of that section can be extended to sometimes
better estimators, we consider the BHHJ estimators
of \citet{BHHJ98, JHHB01, basuetal11}. Methods and
results do extend to general regression setups
but for ease of presentation we consider the i.i.d.~scene,
with $y_1,\ldots,y_n$ from some unknown density $f$
with c.d.f.~$F$.
For a tuning parameter $a>0$, the estimator $\hatt{\btheta}_a$
is defined as the minimizer of the criterion function
$
H_n(a)=\int f(y,\btheta)^{1+a}\,\dd y-(1+1/a)\,n^{-1}\sumin f(y_i,\btheta)^a.
$
These have good properties, affording robustness
without losing much in efficiency, for small values of $a$,
and the ML estimator is the limiting case $a\arr0$.
Theorems in the above mentioned references
give limit distributions for $\rootn(\hatt{\btheta}_a-{\btheta}_{0,a})$,
involving the least false parameter the BHHJ estimator
is aiming for; for $a=0$, this is the minimizer of
the Kullback--Leibler distance $\KL(f,f_{\boldsymbol\theta})$ associated
with ML estimation.

Just as FIC was constructed with ML
estimators as in Section~\ref{subsection:jullumhjort},
formulae can be extended to reach versions of
(\ref{eq:jullumhjortA}) and (\ref{eq:jullumhjortB}),
now instead employing $\hatt{\btheta}_a$ to form
$\hatt\mu_\para=\mu(\hatt{\btheta}_a)$;
details are in \citet{HjortWalker26a}.
The picture broadens: when estimating the focus $\mu$,
there is for each candidate model a continuous class
of estimators $\hatt\mu_{\para,a}=\mu(\hatt{\btheta}_a)$,
with $a$ the tuning parameter, and an associated
$\fric(a)$ curve, for Focused Robust Information Criterion.
The curve's minimizer gives the best tuning parameter, typically with a positive
probability that the method simply points to the ML
estimator, i.e.~$a=0$. The $\fric(a)$ curves
are informative, giving both the best tuning parameters
for each model, and automatic comparison with $\fic_\nonpara$,
the best estimated $\mse$ for the nonparametric
$\hatt\mu_\nonpara=\mu(F_n)$.

\textbf{Figure \ref{figure:egypt}} illustrates the FRIC
to estimate $\mu=F^{-1}(0.75)$, the 0.75 quantile
of the lifelength distribution in Roman Era Egypt,
a century b.C.; see \citet[Ch.~3]{ClaeskensHjort08} for relevant
background information regarding what's behind the $n=141$
lifelengths excerpted from engravings on mummies.
The direct 0.75 sample quantile is $\hatt\mu_\nonpara=48.0$ years,
which we now compare with estimates from
the two-parameter gamma and Weibull models.
\textbf{Figure \ref{figure:egyptA}} displays the $\fric(a)^{1/2}$ curves,
showing that the best tuning parameters for BHHJ estimation
are at 0.464 and 0.807, for these models, and that
the resulting $\hatt\mu_{\para,a}$ estimates are
better than the nonparametric, in that the minima are
below the $\fic_\nonpara^{1/2}$ value.
\textbf{Figure \ref{figure:egyptB}} has the estimates on the y-axis,
as functions of the root-FRIC values. The best estimates
are 45.9 and 45.8 for the gamma and Weibull models,
again both being better than the nonparametric 48.0.
Instructively, that figure also shows that when
using the FIC of Section \ref{section:ficB},
with $a=0$ for the ML estimates, indicated in
the figure with triangles, the nonparametric estimate
would win. So it pays off using a more flexible estimation
scheme when working with parametric models.

As with FIC in other setups, the FRIC scheme delivers
different winners for different focus parameters,
and here, in particular, for different $F^{-1}(q)$ quantiles.

There are several other domains where the FIC approach
towards selecting tuning parameters is suitable.
Consider e.g.~ridge regression, for data
$y_i\sim\N(\bm{x}_i^\tr\bbeta,\sigma^2)$ with parameter dimension
$p$ often bigger than sample size $n$. The usual
ridge approach is to minimize
$\sumin (y_i-\bm{x}_i^\tr\bbeta)^2+\lambda\|\bbeta\|^2$,
perhaps with cross validation involved in finding
one ostensibly good $\lambda$, then to be used for
all further inference. The focused approach developed
in \citet{HelltonHjort18}, called FRIDGE,
selects one fine-tuned $\lambda(x_0)$ when
making predictions for each given individual, with
his or her covariate vector $x_0$.
This makes a clear contrast with cross-validation
for the tuning parameter choice, where one single $\lambda$
is selected for all predictions, regardless what is $x_0$.
By being focused, the FRIDGE tuning can result in achieving
smaller average prediction error.

For another type of fic-ification, for i.i.d.~data $y_1,\ldots,y_n$
symmetrically distributed around an unknown $\mu$, consider
the Huber estimator $\hatt\mu_a$, the minimizer of
$\sumin h_a(y_i-\mu)$, with $h_a(u)$ the Huber loss function,
which is $\half u^2$ for $|u|\le a$ and $a(|u|-\half a)$
for $|u|\ge a$. For $a$ low and $a$ large we retrieve
the sample median and sample mean, respectively.
There is a clear limit distribution
$\rootn(\hatt\mu_a-\mu)\arr_d \N(0,\tau_a^2)$.
The FIC approach here, where there is no bias under the
symmetry assumption, is to construct a
good $\hatt\tau_a$ from the data and then minimize that expression.
This is admittedly a non-trivial exercise, involving
also estimating the density $f$ itself, perhaps via kernel
smoothing, but illustrates the general FIC thinking
in such fine-tuning problems.
See Supplement Section E 
for some details. 
Such constructions are also possible in regression
setups with robust estimators, for
bandwidth selection for classes of density estimators, etc.

\section{POST-SELECTION INFERENCE}
\label{section:postselection}

Once a good model has been selected, the story does not end there yet. Often, one is interested in inference using the selected model. For example, in the Diabetic retinopathy study of Section~\ref{subsection:klein2008}, for strata 2 and 4,
$
p(\bm{x},\bm{z})
=\Pr(Y=1\midd \bm{x},\bm{z})
=H(\beta_0+\beta_1x_1+\beta_2x_2+\gamma_3z_3+\gamma_5z_5),
$
is the selected model for the specified focus parameters. If one would want to use this selected model to produce confidence intervals for the population version of its parameters or for functions thereof,  or to construct hypothesis tests related to these parameters, classical theory does not apply. Indeed, classical textbook results start with phrasing hypotheses using a correctly specified model, then data come into play and the hypothesis test is executed.
This is absolutely not what happens in our case. We started by using the data to select a model. Once we see which parameters appear in that model, we phrase hypotheses only about those parameters (or functions thereof). This is an example of what is called `data snooping'.

The dangers with using classical methods after model selection have long been understood and pointed out \citep[e.g.][]{Hotelling1940, Breiman01}.
When the same data are used twice, the classical (also called `naive') p-values and confidence intervals are too optimistic: there are too many rejections of the null hypotheses because `naively' computed p-values are too small, with a distribution that is not uniform under the null hypothesis and the confidence intervals have a smaller coverage than the classical nominal level would suggest.

\subsection{Post-selection inference under
  a local neighborhood models framework}
Using the local neighborhood assumption, it is clear that the limiting distribution of the estimator $\widehat{\mu}_{\widehat{S}}$ where the set $\widehat{S}$ is the result of a selection procedure, has a complicated form and is no longer normal.
Let us write the estimator after selection in the form of a model averaging estimator
$\hatt\mu_{\widehat{S}}= \sum_{S\in\mathcal{S}} c(S\midd \bD)\hatt\mu_S$,
  where $\mathcal{S}$ represents the prespecified collection of models
  and the data dependent weights $c(S\midd \bD)$ for model selection are
  all zero  except for $c(\widehat{S}\midd \bD)=1$.
  The more general model averaging estimators may use other data-dependent
  weights, typically required to sum to one and with at most a countable
  number of discontinuities.
Extending the Master Theorem I, which specified the limiting
distribution of a focus estimator for a fixed set $S$,
there is a Master Theorem II, which describes the
limiting distribution not only of the finally selected
$\hatt\mu_{\hatt S}$, but of any perhaps complicated mixture estimator
$\hatt\mu^*=\sum_S c_n(S\midd\data)\hatt\mu_S$,
as long as there is convergence of the weights
$c_n(S\midd\data)\arr_d c(S\midd \bD)$.
From \citet[][Theorem 7.1]{ClaeskensHjort08},
\beq
\rootn(\hatt\mu^*-\mu_\true)\arr_d
\sum_{S\in\mathcal{S}} c(S\midd \bD)\Lambda_S =\Lambda_0
+ \bomega^\tr (\bdelta-\hatt{\bdelta}^*),
{\rm with\ }
\hatt{\bdelta}^*=\sum_{S\in\mathcal{S}} c(S\midd \bD) \bG_S \bD.
\label{eq:master2}
\eeq
First, the generality of this limit result ought to
be noted and appreciated. The theorem covers post-selection
cases, for different selection schemes, including the AIC
and versions of FIC, as well as model-averaging estimators.
These correspond to different ways of setting up
classes of the $c_n(S\midd\data)$ weights,
leading in the limit experiment to clear $c(S\midd \bD)$.
The core structure is the behavior of all such estimators
is determined by the cleverness of $\hatt\psi=\bomega^\tr\hatt\delta^*$
to estimate $\psi=\bomega^\tr\delta$.
Second, it is clear that the resulting distributions
might be quite complicated. Each component $\bG_S \bD$
involved in the $\hatt\delta^*(\bf D)$
is a clear normal, with bias and variance, but the
resulting $\hatt\psi$ is then a nonlinear mixture
of these normals, depending on the position of $\bdelta$
in its parameter space.
See \citet[][Section 7.5]{ClaeskensHjort08}
for a two-stage simulation approach
for the construction of confidence intervals.

\subsection{Marginal post-selection inference for a saturated true model}
Another approach to valid post-selection inference for normal linear models is provided by the conservative method of \cite{BerkBrownBujaZhangZhao13}. The true model in this approach is a saturated model where
$\bm{Y}\sim \N_n(\bmu,\sigma^2\bm{I}_n)$ for an unspecified vector $\bmu$. A selection (it does not need to be specified which method is used) takes place in a collection of linear models of the form $\bm{Y}=\bm{X}\bbeta+\bvarepsilon$. This implies that the pseudo-true value, or the target of inference, is
$\rm{E}(\widehat{\beta}_j)$ for $j\in\widehat{S}$.
A new constant is computed to replace the normal (or t) quantile in the classical confidence interval construction such that the coverage is guaranteed to be at least the nominal value $1-\alpha$, regardless of the selection method. This constant depends on the covariate matrix $\bm{X}$, the set of models $\mathcal{S}$ to select from and the probability of a type I error $\alpha$.
This construction has been extended to allow for heteroscedastic data and for application to binary regression models by \cite{BachocPreinerstorferSteinberger2016}.

\subsection{Conditional post-selection inference}

When information about the model selection procedure is known, as is the case when an information criterion is used for the selection, another approach to arrive at valid inference after selection is to condition on the event of the selection and to use the conditional distribution for inference.  Such selective inference approaches have been described for selection by the AIC \citep{CharkhiClaeskens18}, for likelihood and selection methods based on testing \citep{RugamerGreven18}, by means of adjusted $R^2$ \citep{PirenneClaeskens24} and for regularized estimation methods using the lasso method \citep[][among others]{LeeSunSunTaylor2016, TianTaylor2017}.

Recent methods add a user-controled amount of randomization during the selection in order to use the full set of data for both selection and inference \citep{TianTaylor2018, PanigrahiTaylor2023, PanigrahiFryTaylor2024, HuangPirennePanigrahiClaeskens2025}, to name just a few references.
The advantage of such a selective inference approach is that the full sample can be used; in this sense the method is preferred above sample splitting where only part of the data are used for selection, hence less accurate selection, and where the other part of the data are used for inference, resulting in a loss of power due to the smaller sample size. As compared to the marginal approach, selective inference uses information about the selection in the conditioning event and can henceforth deliver confidence intervals that are either exact or less conservative as with a marginal approach.

\subsection{Bagging and bootstrapping}

\citet{Efron14} studies the use of bootstrap smoothing, also known as `bagging' to incorporate the effects of model selection.
In this approach multiple bootstrap samples are taken either directly from the data (nonparametric bootstrap) or from a parametric model (parametric model). With each bootstrap sample the model selection is performed, resulting in one bootstrap estimator of a focus parameter $\mu$ per bootstrap sample. The bagging estimator takes the sample average of the bootstrap estimators over a number $B$ (Efron uses 4000 in the paper's example) different such bootstrap samples. In a comment to this paper \cite{Hjort14} considered parametric resampling from a wide model and studied the bagging estimator under a local misspecification scenario, leading to a generalisation of the master Theorem II to apply to bagging estimators. See \cite{Hjort14} for more details.

\section{OTHER MODEL CLASSES, ESTIMATORS AND LOSS FUNCTIONS}
\label{section:othermethods}

In the previous sections the data were assumed to be independent,
we mainly used ML estimators and the aim was to minimize
the estimator's mean squared error.
In this section we discuss three divergences from the basic setup:
using data that are not independent, a main example is the context
of time series; using other estimators than the ML; and
using loss functions that are different from squared error loss,
hence stepping away from the estimated mean squared error
as the objective to be minimized.

\subsection{Time series, spatial and longitudinal or panel data}
\label{subsec:timeseries-etc}

One extension from the basic setup of the previous sections is focused selection for data that are not all independent, with main applications in econometrics, spatial statistics, and biostatistics and medicine.

As in \citet{Lohmeyer2019}, consider a vector autoregressive (VAR) model with a local-to-zero parametrization where there are $k$-dimensional time series organized in a vector triangular array $\{\{\bm{Y}_{T,t}\}_{t=\infty}^\infty\}_{T=1}^\infty$ such that with $k\times k$ coefficient matrices $\bm{B}_1,\ldots,\bm{B}_{p_1}$ and $\bm{\Delta}_1,\ldots,\bm{\Delta}_{p_2}$, the VAR model is
$
\bm{Y}_{T,t}=\sum_{j=1}^{p_1} \bm{B}_j \bm{Y}_{T,t-j} +
\sum_{j=t}^{p_2} 
T^{-1/2}\bm{\Delta}_j\bm{Y}_{T,t-p_1-j}+\bm{\bepsilon}_t,
$
for an independent and identically distributed $k$-variate sequence
$\{\bepsilon_t\}_{t=1}^\infty$ with zero mean and positive definite covariance matrix.
An observation at time $t$ depends on the previous observations up to $p_1+p_2$ time-units back in time.
From a joint asymptotic normality result for the estimators of the elements of the coefficient matrices and of the covariance matrix of the errors $\bepsilon_t$, \citet{Lohmeyer2019} obtain the asymptotic normality and the mean squared error of focus estimators to define a FIC. Of particular importance are the impulse responses, and the accumulated responses. For examples of simultaneous focused selection of the autoregressive order and regression variables, see  \citet{Claeskensetal07}.
Focused selection of focus quantities from generalized linear models with time series data is developed in \citet{PandhareRamanathan20}. Once the limiting distribution of the focus estimators in the different models have been obtained, the FIC can be readily constructed from the mean squared error expressions.
\citet{HermansenHjortJullum15} compared parametric and nonparametric models for stationary time series using the FIC.

Also spatial effects cause dependence among the observations.
\citet{Pandhare25arxiv} use a spatial lag model of the form $(\bm{I}_n - \rho \bm{W}) \bm{Y} = \bm{X\beta} + \bm{\varepsilon}$, where $\bm{W}$ is a $n\times n$ adjacency matrix. The coefficient vector is estimated for known $\rho$ via two regression models, one with $\bm{Y}$ as the response and the other with response $\bm{WY}$. Several focus quantities  are of interest, including the spatial spillover effect $\log|\bm{I}_n- \rho \bm{W}|$ for which they developed an FIC.

Longitudinal or panel data display dependence too. Focused selection for count responses in panel data is developed by \citet{WangLiSun15}. For focused selection in dynamic panel models, see \citet{ChangDiTraglia18}. \citet{YinLiuLin21} considered selection in panel models with a multifactor error structure.
Only allowing focused selection within the fixed effects of a longitudinal model, see \citet{HuChengZeng21} and \citet{YangPengZouLiang17}. More general focuses also allowed to depend on the variance components of a linear mixed model, were studied in \cite{CunenWalloeHjort20}.

\subsection{Quasi-likelihood and quantile regression models}

The ML estimators are a common choice for focus estimation,
though other estimation methods have been applied in combination
with a focused information criterion.

Quasi-likelihood estimators only make assumptions about moments of the response vector, not about its complete distribution. When the mean of the response is correctly specified, unbiased estimating equations can be constructed. For focused selection,
\citet{RamanathanPandhare21} study quasi-likelihood estimating functions that are locally biased in the context of logistic regression time series models. 
The focused selection taking the bias in the estimating equations into account is found to outperform the version that assumes unbiasedness.
Quasi-likelihood estimators formed the starting point for focused model selection in
generalized additive partial linear models \citep{ZhangLiang11};
for focuses in high-dimensional generalized linear models  \citep{PandhareRamanathan23};
and for models for recurrent event data subject to left-, right-, and intermittent-censoring
\citep{StoltenbergHjort21}.
A focused information criterion has been developed based on generalized estimating equations (GEE) for longitudinal data \citep{YangPengZouLiang17}.
For generalized method of moment estimators the asymptotic normality of the estimators formed the basis for a mean squared error-based focused information criteria, see
\citet{ChangDiTraglia18}, applied to selection of focuses for dynamic panel models.

The estimation via quantile regression models requires specific care because the `check' function 
$\tau I(u\ge0)+(\tau-1)uI(u<0)$  used to estimate a $\tau$-quantile is non-differentiable at the origin. Some examples of focuses for the FIC for quantile regression include the focused selection of the fuel price elasticity of transport demand \citet{BehlDetteFrondelVance19} and of a minimum effective dose in phase II clinical trials \citep{BehlClaeskensDette2014}. For the related focused search
of a benchmark dose using ML estimators, see \citet{PeñaWuPiegorschWestAn17}.
A focused selection has been studied in quantile regression models for right-censored data with parametric models by \citet{DuZhangXie17} as well as with partially linear models
\citep{SunSunLuZhiLi17}.

\citet{Sueishi13} constructed a FIC for selection among
generalized empirical likelihood estimators which extend the empirical likelihood estimators and the exponential tilting estimators and applied it to selection in linear instrumental variable models.

\citet{ZhangLi24} encompass several focused selection methods by defining the estimators as the maximizers of some objective function. Under some assumptions (including the existence of third derivatives of the objective function with respect to the model parameters) they obtain the asymptotic normality of the estimators, from which a mean squared error expression is obtained and estimated to form the FIC.

Outlier-robustness of estimators can be dealt with via M-estimation methods, see also Section~\ref{section:fic-finetuning}. 
\citet{DuZhangXie18} develop an outlier-robust FIC for selection in a collection of linear models.
\citet{PandhareRamanathan2020} constructed a robust FIC using M-estimators in autoregressive models.
\citet{PandhareRamanathan23} work with high-dimensional generalized linear models and develop an asymptotic normality result for desparsified M-estimators using the local misspecification framework. From this a robust FIC is obtained.

\subsection{Minimizing other risk functions}

While the mean squared error minimization is the objective for the vast majority of the focused selection literature, minimizing other than squared error loss could be relevant too.
After the specification of the focus, the main steps for the FIC are to construct and estimate the relevant loss function for the estimators for each model and select the model for which the estimator has the smallest estimated risk.

An asymmetric loss function such as the linex (linear exponential)
loss might be needed when, for example, underestimation might
be more problematic than overestimation, such as when estimating
high water levels for flood predictions.
The linex loss function \citep{Varian75, Zellner86}
takes the form  $L(u)=\exp(cu)-cu-1$ where $c>0$ when overestimation
is a bigger concern than underestimation and $c<0$
in the opposite case. When $c$ is close to zero,
the linex loss is close to the squared error loss.

For a focus parameter $\mu$ with estimator $\widehat{\mu}_S$
in candidate model $S$, the linex loss is
$
L_{n,S}=\exp \{c\rootn(\widehat{\mu}_S-\mu_\true)\}
   - c\rootn(\widehat{\mu}_S-\mu_\true)- 1.
$
From the master theorem I in expression (\ref{eq:master1}),
there is a clear limit in distribution variable
$L_S= \exp(c\Lambda_S) - c\Lambda_S - 1$,
from which the limiting risk $\E(L_S)$ is obtained.
Plugging in estimators as before, we arrive at a FIC that minimizes
the linex risk for the focus estimator;
see the Supplement Section~F 
for details.
See also \cite{ClaeskensHjort08b} for constructing
an AFIC for weighted linex loss function over several focus
parameters.

For binary outcomes it might be relevant to consider the probability of misclassifying a new observation as the risk function. When predicting a new binary response value $Y_{\rm new}$, e.g.~to decide on whether or not a new patient is diagnosed with a particular medical problem, there is independence between the new observation and the sample of data.
For each of a given set of models, the response value can be predicted using that model's estimators. For example, in a logistic regression model with
$\Pr(Y=1\midd \bm{x})=H(\bm{x}^\tr\bbeta)$ one may decide to classify the estimated response as having value 1 when the estimated probability is larger than 0.5, and classify it as zero otherwise.
The best focused estimator for this prediction according to  minimizing the misclassification probability is that one for which an estimated version of
$r(S) = \Pr(Y_{\rm new}=1) \Pr(\widehat{Y}_S=0) +  \Pr(Y_{\rm new}=0) \Pr(\widehat{Y}_S=1)$
is minimal. See \citet{ClaeskensCrouxVanKerckhoven06} for further details.

Another similar loss function of the all-or-nothing type is the following,
in the generic setup with $\hatt\mu$ estimating $\mu$;
\beq
L^*(\mu,\hatt\mu)=1 {\rm\ if\ } |\rootn(\hatt\mu-\mu)|>\eps,\,\,
  0 {\rm\ if\ } |\rootn(\hatt\mu-\mu)|\le\eps.
\label{eq:zeroone}
\eeq
We refer to the Supplement Section~G 
for more details regarding the construction
of a FIC based on this loss function.
In \citet{ClaeskensCrouxVanKerckhoven06} one can also find the FIC for using $L_p$ loss in general with the case of $p=2$ corresponding to using the mean squared error and $p=1$ to using the mean absolute error.

\section{APPLICATION DOMAINS, RECENT DEVELOPMENTS, FUTURE DIRECTIONS}
\label{section:recent}

We briefly mentioned various domains of FIC application
in the introduction, with details to be found
in the references pointed to. Below are some further
such instances, where the FIC, with its variants,
has been demonstrated to be beneficial for constructing
the best estimators for given purposes. 
The cited references may be consulted for more details.

\cite{Hansen05} recommended an investigation of the use
of FIC for econometric model selection; a lot of progress
has been made since. Besides the developments in time series analysis,
see Section~\ref{subsec:timeseries-etc}, other FIC work
for economical data includes
\cite{BehlDetteFrondelVance19, BehlDetteFrondelTauchmann13, BehlDetteFrondelTauchmann12, Brownlees08, DiTraglia16, Klimenka19, WangHH2019},
without attempting to be complete here.

Evidently, in medical practices patients may benefit from
a personalized treatment. It is therefore no surprise
that focused model searches have been used in medical contexts.
\cite{YangLiuLiang2015} and \cite{HelltonHjort18}
use focused selection to aid with an individualized diagnosis. 
Focusing on the area under the curve in a receiver operating
characteristic study, \citet{YangHuangQin17}
use FIC to improve the diagnostic accuracy of a test
with a hearing device. \citet{HjortClaeskens06, JullumHjort19}
address focused selection for survival data in a biomedical contexts,
and \citet{StoltenbergHjort21} develop
a FIC for recurrent event data.
In \citet{ChristensenHjort26} a FIC is constructed for
optimal estimation of the LD50 parameter in bioassays,
the Lethal Dose point for the dose level $x_0$ at which
there are 50-50 chances for survival or death, with
many further applications. 

For applications in biology and ecology, see \citet{ClaeskensCunenHjort19}. 
\citet{HermansenHjortKjesbu16} used FIC to study marine science
time series data to investigate fish quality.
\citet{CunenWalloeHjort20, CunenWalloeKonishiHjort21}
addressed specific focused questions about populations
of Minke whales, with consequences for international
whaling politics. Specific questions in political and social
sciences benefit from an FIC search too.
\citet{Haug19} used FIC for the estimation of the
probability that an armed conflict escalates using Markov chain data,
while \citet{CunenHjortNygaard20} used series of battle-deaths data
for focused searches regarding the median number of battle deaths.
\cite{Hjort26} suggests an FIC use in meteorology and climate sciences. 

In the current era of big data, the high-dimensionality
of the data comes with additional challenges, also for
focused variable selection.
When regularization methods are used for estimation,
the statistical aspects are
more challenging due to the shrinkage bias caused by
the regularization, and for regularization methods
that perform selection (as is the case with an $\ell_1$-penalty)
the effects of the selection should be taken into account as well.
\cite{PircalabeluClaeskensJahfariWaldorp15} perform
focused selections in graphical models that were used
for fMRI data and for which the number of nodes
in the graph exceeds the sample size. To avoid the effects
of the selection due to the regularization,
local quadratic approximations to several penalty functions
(including that for adaptive lasso, hard thresholding,
and  the smoothly clipped absolute deviation)
were used to derive expressions for the FIC.
\citet{GueuningClaeskens18} made a distinction
between FIC expressions for low-dimensional and
high-dimensional submodels. For the latter case
a desparsified version of the $\ell_1$ regularized
estimators has been used that undoes the selection
and has an asymptotic normal distribution.
\cite{PandhareRamanathan23} also use desparsified
estimators for a focused selection among outlier-robust
M-estimators in high-dimensional generalized linear models,
for which they also study the influence functions.
We also point to the copula-FIC methods of \citet{KoHjort19}
as an instance of using FIC to aid modelling
of complex dependences in both moderate and higher dimensions. 

Most of the machine learning methods (such as boosting
and its variants, neural networks etc.)~are intrinsically
very high-dimensional. Several such methods are reported
to perform quite well in terms of out-of-sample predictions.
It would be worthwhile to study how a focus can be
incorporated in such methods to arrive at even better predictions.
Many interesting research directions use high-dimensional
weather data for nowcasting, this is a prediction
at the current time,  
which is important in many domains, from health care
to insurance to tourism to just name a few.
Focused searches would be utmost helpful here,
to predict the number of hospital beds during the flu season,
the number of insurance claims after a storm,
or the number of hotel accommodations in the summer peak period.

The FIC should find more uses in classification and discriminant
analysis applications. Deciding whether a new object
with measurement vector $x$ is of type 1 or type 2
depends crucially on the density ratio $f_2(x)/f_1(x)$, say,
so FIC could be constructed for selecting the best models 
for $\mu(x)=\log f_2(x)-\log f_1(x)$, with highest importance
given to the $x$ for which the $\mu(x)$ are close to zero.
For some types of models, this translates into using FIC
to aid model selection in logistic regressions,
precisely for the most difficult cases where the
computed probabilities are close to 0.50. 

The era of `combining information sources' is thriving
and growing. To answer the perhaps big question $Q$,
researchers identify components $C_1,\ldots,C_k$, say,
extracting relevant partial information, and fuse the pieces
into tentative answers $A$. A general II-CC-FF paradigm
for such general schemes is proposed and broadly illustrated
in \citet{CunenHjort21}, involving Independent Inspection,
Confidence Conversion, and Focused Fusion;
the more familiar meta-analysis methods are special case.
The present point is that extra layers of FIC might come into
the play, for both II and FF. Such methods might need
to be worked out on a case by case basis. 

Developing FIC methodology further, in various directions, 
requires of course mathematical efforts for solving
new pertinent questions, regarding approximation quality
for risk functions, performance of competing model averaging
schemes, accurate confidence intervals, etc., in a widening
list of setups. Of relevance here is also the fact
that different FIC scores, being themselves estimators
with different disributions and accuracies, might be
polished further. In \citet{CunenHjort20} confidence distributions
are constructed for the FIC scores, giving insights
into the accuracy of the selections schemes and leading
also to new relatives, like the median-FIC selector.
The FIC methods surveyed in our article have essentially
been developed for setups with `inner parameters',
not close to hard boundaries, for which more delicate
approximations become necessary. \citet{Hjort94} is a case
in point, aiming to distingush the t from the normal;
the latter is the special case of $\df=\infty$,
not an inner point among the normals.
More work would be needed to build better
FIC for variance components models, for example.

\section{CONCLUDING REMARKS}
\label{section:concluding}

We have reviewed good portions of the growing FIC literature,
seen FIC and AFIC in action, and pointed to yet further
issues and directions. 

{\textit{Some practical concluding recommendations}.
  The FIC allows quite specific model choices. A first advice is to set a good research question. Which quantity is of importance? This will define the focus. Which model class is useful for the data at hand to estimate that quantity? This will lead to a set of models.
While these questions would also be asked when applying any other 
model selection method, the FIC explicitly uses the defined focus in its construction to point towards the best model in the specified set of models. 
When there is a focus, for example estimation of the 0.9 quantile of the distribution of $Y$ given the covariate $x$, but no specific covariate value $x_0$ is of interest, the AFIC would be the preferred choice. The choice between a fixed or local wide model depends on the assumptions the researcher is willing to make. When domain knowledge or prior experience suggests the use of a certain parametric distribution, a local wide model would be convenient, whereas in situations with more doubt, a nonparametric wide model might be more appropriate.

A brief list of further remarks is as follows.

{\it A. FIC approximations.}
When constructing the FIC and AFIC, in different
frameworks, we have needed to rely on certain approximations,
both for the local and the fixed wide model setups,
involving limit distributions, Taylor expansions, etc.
When $n$ is small or moderate, compared to the model dimension,
modifications may be in order; see some discussion
of this in the JASA discussion to
the \citet{HjortClaeskens03a, ClaeskensHjort03} papers,
and the rejoinder. It is comforting, though,
to work through the algebra of the linear regression models,
and learn that the FIC approximations are then exact,
for focus parameters being linear combinations of the mean parameters.
This is true for both lines of FIC approximations,
the local and the fixed wide model;
in particular, the two lines of FIC constructions
give identical formulae here.

{\it B. The role of the sample size.}
With increasing sample size $n$, the FIC will tend
to select more complex models, which matches statistical
intuition. In the framework of Section \ref{section:ficB},
the risks are of the form $b_j^2+v_\para/n$.
If none of the candidate models have $b_j=0$, i.e.~zero
bias for the focus parameter, the FIC will with probability
tending to 1 select the widest model. If the parametric
model under consideration has $b=0$, however,
there is still a chance that FIC prefers the nonparametric.
Indeed this probability can under some conditions
be seen to tend to $\Pr(\chi^2_1>2)=0.157$;
see \citet[Section 4]{JullumHjort17}, a special case
of the limiting probability $\Pr(\chi^2_\df>2\cdot\df)$
pointed to in Section \ref{subsection:catfic}.

{\it C. Balancing the narrow and the wide model.}
We learn from Master Theorems I and II,
see eq.~(\ref{eq:master1}) and (\ref{eq:master2}),
that crucial roles are played by $\tau_0^2$,
the narrow model variance present for all methods,
and the maximal extra variance $\bomega^\tr \bQ \bomega$
associated with the wide model. The variance ratio
$\rho=\tau_0^2/(\tau_0^2 + \bomega^\tr \bQ \bomega)$
can be 
estimated from data. If $\rho$ is high,
different selection schemes will behave similarly;
if on the other hand $\rho$ is small, there is more to
win by using clever selection and model averaging schemes,
and, in particular, by using FIC.

{\it D. Performance studies.}
Model selection and model averaging schemes are complex,
with no clear winners; details of performance depend
on the framework, the focus, the list of candidate models,
the sample size, and the perhaps many parameters.
Master Theorems I and II make clear that core issues
are determined by the behavior and risk functions
in the limit experiment, however. Thus, when stripping away
extraneous secondary aspects, matters boil down to studying
how well the estimator $\hatt\psi=\sum_S c(S\midd \bD) \bG_S \bD$
does for estimating the linear combination $\psi=\bomega^\tr \bdelta$
in the normal mean problem where $\bD\sim\N_q(\bdelta,\bQ)$
and $\bQ$ is known. Examining behavior, for different
strategies, could therefore start there, studying risk
functions of the different $\hatt\psi$, for different
linear combinations and positions in the parameter space.

{\it E. Good candidate models.}
We have seen that the FIC with its various variants aims
at and will often succeed in finding the best among
the given list of candidate models. The FIC is not a panacea
that the winner is perfect, 
and there is no contradiction
if another model could be found to be even better.
Efforts should be spent in setting up and also
limiting the list of candidate models. As the broad
discussion of \citet{McCullagh02} reminds us, many
models ought to be screened away, on good statistical prior grounds,
even if they make mathematical sense. This FIC warning
is also pertinent regarding the choice of the `fixed wide model'
in the broader framework of Section \ref{section:ficB}.
A case in point is \citet{CunenWalloeHjort20, CunenWalloeKonishiHjort21},
where an international scientific committee spent lengthy
efforts on agreeing on the particular wide model,
from a class of linear mixed models, and then,
a fortiori, on what the FIC then delivered,
with international political consequences for whale quotas. 

{\it F. Links to Bayes.}
Several Bayesian versions of the FIC are developed in 
\citet{RoyLesaffre26}, by estimating the mse
using the posterior distribution of the focus in question.
In their application they focus on the average body
mass index of children at one year of age and select
among several longitudinal data models.
Future developments of FIC in the Bayesian context
for other model classes would be worthwhile.

\section*{DISCLOSURE STATEMENT}
The authors are not aware of any affiliations, memberships, funding,
or financial holdings that might be perceived
as affecting the objectivity of this review.

\section*{ACKNOWLEDGMENTS}
G.C.~acknowledges support from project C16/20/002 of the Research Fund KU Leuven, Belgium and from project ASTeRISK Research Foundation Flanders [grant number 40007517] of the Excellence of Science (EOS)
program (FWO-FNRS, Belgium).
N.L.H.~has appreciated generous funding from the Norwegian Research Council
for leading the five-year project
{\it FocuStat: Focused Statistical Inference with Complex Data}.

\bibliographystyle{ar-style1} 
\bibliography{gerdanils_bibliography_9July2026.bib}

\end{document}